%% file: report.tex
\newcommand{\numColumns}{1}           
\newcommand{\isUseAbstract}{1}        
\newcommand{\isUseTableOfContent}{0}  
\newcommand{\isUseThanks}{0}          
\newcommand{\isUseThightHeaders}{0}   
\newcommand{\dateMode}{1}             
\newcommand{\myBibliographyStyle}{0}  
\newcommand{\myBibliographyFile}{refs.bib}

\newcommand{\myDocumentType}{Article}

\newcommand{\myTitle}{ROPE: A Novel Method for Real-Time Phase Estimation of Complex Biological Rhythms}

\newcommand{\myAuthor}{%
  \styleAuthor Antonio Spallone$^{1,2}$,
  \styleAuthor Marco Coraggio$^{3}$,
  \styleAuthor Francesco De Lellis$^{1}$,
  \styleAuthor Mario di Bernardo$^{1,3}$%
}

\newcommand{\myAffiliation}{%
  \styleAffiliation
  $^{1}$University of Naples Federico II, Naples, Italy\quad
  $^{2}$Polytechnic University of Bari, Bari, Italy\quad
  $^{3}$Scuola Superiore Meridionale, Naples, Italy%
}
\newcommand{\myAbstract}{%
  Accurate phase estimation---the process of assigning phase values between $0$ and $2\pi$ to repetitive or periodic signals---is a cornerstone in the analysis of oscillatory signals across diverse fields, from neuroscience to robotics, where it is fundamental, e.g., to understanding coordination in neural networks, cardiorespiratory coupling, and human-robot interaction.
However, existing methods are often limited to offline processing and/or constrained to one-dimensional signals.
In this paper, we introduce ROPE, which, to the best of our knowledge, is the first phase-estimation algorithm capable of (i) handling signals of arbitrary dimension and (ii) operating in real-time, with minimal error.
ROPE identifies repetitions within the signal to segment it into (pseudo-)periods and assigns phase values by performing efficient, tractable searches over previous signal segments.
We extensively validate the algorithm on a variety of signal types, including trajectories from chaotic dynamical systems, human motion-capture data, and electrocardiographic recordings.
Our results demonstrate that ROPE is robust against noise and signal drift, and achieves significantly superior performance compared to state-of-the-art phase estimation methods.
This advancement enables real-time analysis of complex biological rhythms, opening new pathways, for example, for early diagnosis of pathological rhythm disruptions and developing rhythm-based therapeutic interventions in neurological and cardiovascular disorders.%
}

\ifnum\numColumns=1
  \documentclass[onecolumn]{article}
\else\ifnum\numColumns=2
  \documentclass[twocolumn]{article}
\else
  \GenericError{}{value of \protect\numColumns\space is not acceptable}{}{}
\fi\fi

\input{preamble_report.tex}

\begin{document}

\maketitle
\thispagestyle{firstpage}  
\ifnum\isUseAbstract=1
\begin{abstract}
  \noindent \normalsize {\color{colorAccent}{\textsf{\textbf{Abstract. \;}}}}\myAbstract
\end{abstract}
\fi
\vspace{3ex}

\ifnum\isUseTableOfContent=1
  \tableofcontents
\fi

\section{Introduction}

Rhythmic and repetitive signals are pervasive across biological and engineered systems, with particularly rich examples found in human cardiovascular and respiratory systems \cite{Schfer1998,Lotri2000}, neural network activity \cite{Rosenblum2021}, human and animal locomotion patterns \cite{Kobayashi2016}, and adaptive robotics \cite{Maeda2017}.
In these domains, understanding the temporal organization and coordination between oscillatory components requires estimating the (possibly time-varying) periods of signals and assessing synchronized activity across multiple subsystems, which is typically achieved through \emph{phase estimation}, the process of assigning instantaneous phase values (e.g., between $0$ and $2\pi$) to each instant of a dynamic signal.

This phase information reveals fundamental coordination mechanisms across biological scales. In neuroscience, phase dynamics underlie theories of neural coordination, selective attention, and inter-regional communication through coherence \cite{Fries2005,Canolty2010,Higgs2017}, with phase disruptions linked to neurological and psychiatric disorders. In cardiology, phase response curves and phase synchronization measures quantify cardiorespiratory coupling and autonomic regulation \cite{Kralemann2013,Bartsch2015}, providing clinical insights into cardiovascular health. For human motion analysis, continuous phase estimation enables detailed characterization of gait cycles \cite{Manzoori2023} and coordination dynamics in joint actions \cite{Schmidt1990,Varlet2011}, informing rehabilitation and motor learning strategies. 
Similarly, robotics and human-robot interaction have leveraged phase-aware control frameworks to achieve adaptive behaviors in collaborative tasks \cite{Maeda2017,Saveriano2023,grotta2024learningbaseda}, creating more intuitive and responsive human-machine interfaces.
In the literature, phase estimation of real-world signals is most often achieved by using the Hilbert transform.
This method is based on the concept of the \emph{analytic signal}, introduced by Gabor in \cite{Gabor1946}, which represents a real-valued signal as a complex-valued function from which instantaneous phase can be extracted. 
This approach has been extensively applied across neuroscience, physiology, and biomedical engineering \cite{LeVanQuyen2001, Harrison2021, Peng2019, Hu2008}, becoming a standard tool for analyzing neural oscillations, respiratory patterns, and cardiac rhythms.
However, the Hilbert transform presents several fundamental limitations that constrain its effectiveness for biological and physiological signal analysis.
The method is inherently non-causal and exhibits sensitivity to signal bandwidth and noise characteristics \cite{Rosenblum2021,Matsuki2023}, making it unsuitable for real-time applications such as closed-loop medical devices, brain-computer interfaces, and adaptive rehabilitation systems where immediate phase feedback is essential.
Additionally, the standard Hilbert transform operates exclusively on one-dimensional signals, necessitating dimensionality reduction techniques when applied to the multidimensional data typical of biological systems---such as multi-channel neural recordings, three-dimensional movement trajectories, or spatially distributed cardiac electrical activity. 
This dimensional reduction often discards critical information about the complex spatial-temporal relationships that characterize biological coordination.

Several studies have addressed real-time phase estimation using both geometric and model-based approaches to overcome the limitations of traditional methods. 
Geometric methods, such as the \emph{continuous relative phase} approach by Varlet and Richardson \cite{Varlet2011} or the \emph{state-space method} by M\"ortl et al. \cite{Mrtl2012}, provide phase representations of human motion trajectories while accommodating for frequency modulations and movement variability typical of biological systems.
These methods typically operate on one-dimensional data, requiring projection of multidimensional movement---such as gait, reaching, and interpersonal synchronization---onto a single dimension for real-time analysis.
However, this dimensionality reduction can obscure critical coordination patterns in multi-channel data, such as recordings from multiple brain regions or distributed muscle activity. Moreover, geometric methods often fail to generalize across signal types or handle noise and abrupt changes, such as those in severe cardiac arrhythmias, complex multi-frequency neural oscillations and electrode interference, or artifacts in motion capture data.

Model-based techniques, conversely, estimate phase by fitting mathematical models of oscillatory behavior to observed biological signals, inferring phase evolution from assumptions about the underlying generative processes.
Nonetheless these approaches may struggle with biological oscillators that shift between different dynamical regimes, in response to physiological or pathological changes, and extensive parameter tuning often limits practical applicability.
Examples of this approach include early work on phase reconstruction in chaotic systems \cite{Yalnkaya1997} and more systematic methods for retrieving phase and coupling functions from time series of coupled oscillators \cite{Kralemann2008}.
More recent efforts have focused on simultaneous real-time estimation of phase and amplitude, leveraging synchronization principles and biophysical modeling.
Notably, Rosenblum et al. \cite{Rosenblum2021} developed a method based on causal filtering and entrainment theory, designed for neural data analysis.
Similarly, Wodeyar et al. \cite{Wodeyar2021} introduced a state-space modeling framework that tracks phase as a latent variable, demonstrating improved robustness when processing noisy, multi-frequency biological signals typical of neural recordings.

While these advances represent significant progress toward real-time biological phase estimation, broader applicability is hindered by fundamental limitations.
Most critically, these techniques (i) are generally limited to one-dimensional signals, (ii) often assume relatively smooth and stable temporal dynamics, and/or (iii) are tailored to specific application domains or conditions, limiting generalizability.
These factors make them unsuitable for the complex, multidimensional patterns typical of many biological systems, such as whole-body movement, multi-channel neural activity, or spatially distributed cardiac electrical patterns.

In this work, we address these key limitations by introducing ROPE (Recursive Online Phase Estimation), an algorithm specifically designed to enable real-time phase estimation from multidimensional pseudo-periodic biological signals without requiring dimensionality reduction or domain-specific assumptions. 
ROPE operates by identifying repetitive patterns and efficiently searching past signal segments to determine the current phase; this approach is naturally suited to the variable, multidimensional nature of biological oscillations.
We extensively validate the algorithm across three representative categories of biological signals that span different complexity levels and noise characteristics: chaotic dynamics using the Rössler attractor system (modeling complex neural oscillations), diverse human movement patterns captured through high-precision motion tracking (representing natural motor variability), and electrocardiographic recordings from clinical databases (exemplifying noisy physiological signals). 
Our comprehensive evaluation demonstrates that ROPE significantly outperforms established methods, including the Hilbert transform and PCA-based approaches, achieving superior phase estimation accuracy across all signal types while maintaining the real-time performance essential for biological applications.
Notably, ROPE maintains robust performance even in challenging scenarios where classical methods fail due to signal complexity, noise, or multidimensional structure.

To facilitate adoption and further research, we have made the ROPE algorithm freely available online.%
\footnote{\url{https://github.com/SINCROgroup/recursive-online-phase-estimator}.}

\section{Mathematical notation and preliminaries}
\label{sec:methods}

We start by providing the notation necessary for the formal description of our phase estimation algorithm which will be presented in Section \ref{Sec:algorithm} and introduce definitions that will be used throughout the rest of the paper.

\subsection{Notation and definitions}
\label{sec:notation}

In what follows,
$a \bmod b$ denotes the modulo operation (the remainder of $a / b$).
$\mathsf{i} = \sqrt{-1}$ is the imaginary unit.
$\BB{S} \coloneqq \BB{R}/2\pi\BB{Z}$ is the 1-sphere.
$\angle$ is the phase of a complex number.
$\odot$ is the Hadamard (element-wise) product,
$\circledslash$ is the Hadamard division, and
$^{\circ n}$ is the Hadamard power $n$.
$\vec{1}_n$ is the column vector of $n$ ones.
$\norm{\cdot}_p$ denotes the $p$-norm.
$\ON{dist}_{\R{F}}(\gamma_1, \gamma_2)$ is the Fréchet distance \cite{EiterMannila1994} between curves $\gamma_1$, $\gamma_2$.
Given $f : \C{A} \rightarrow \C{B}$, and $\C{A}^- \subset \C{A}$, $\ON{restr}_{\C{A}^-}(f) : \C{A}^- \to \C{B}$ is the \emph{restriction} of $f$ to $\C{A}^-$.
A reference frame is denoted by $\C{R} = (\vec{o}, \vec{u}_1, \dots, \vec{u}_d)$, where $\vec{o} \in \BB{R}^d$ is the origin of the frame and $\vec{u}_1, \dots, \vec{u}_d \in \BB{R}^d$ are orthonormal vectors giving the orientation of the frame.

By \emph{signal} we mean a function having a subset of $\BB{R}$ or $\BB{Z}$ as a domain (continuous or discrete time, respectively). 
The \emph{continuous-time generalized position} is defined as the $d$-dimensional signal $\vec{p}^{\R{c}} : \BB{R} \rightarrow \BB{R}^d$, describing the complex oscillatory pattern of interest;%
\footnote{The methodology presented in this paper can also be applied to patterns that are not necessarily associated only to periodic kinematic motion but to general oscillatory patterns in biology, physiology or other domain of interest (EEG, ECG, etc.).}
the \emph{discrete-time generalized position} $\vec{p} : \BB{Z} \rightarrow \BB{R}^d$ is obtained by sampling $\vec{p}^{\R{c}}$, with sampling time $T_{\R{s}}$, so that $\vec{p}(k) = \vec{p}^{\R{c}}(k T_{\R{s}})$ (see Figure \ref{fig:signals_in_time}).
The \emph{continuous-time generalized velocity} is defined as $\vec{v}^\R{c}(t) \coloneqq \frac{\R{d} \vec{p}^\R{c}}{\R{d}t}(t)$;
the \emph{discrete-time generalized velocity} is obtained via sampling, similarly to $\vec{p}$, and is defined as $\vec{v}(k) \coloneqq \frac{\R{d} \vec{p}^\R{c}}{\R{d} t}(k T_{\R{s}})$.
The \emph{continuous-time generalized kinematics} is defined as the stack of the continuous-time generalized positions and velocities, i.e., $\vec{m}^\R{c}(t) = [(\vec{p}^\R{c}(t))\T \ \ (\vec{v}^\R{c}(t))\T]\T$. 
The \emph{discrete-time generalized kinematics} is given by $\vec{m}(k) = [\vec{p}\T(k) \ \ \vec{v}\T(k)]\T$.
Finally, a \emph{recording}  is defined as a matrix whose columns are subsequent samples of discrete generalized positions, velocities, or kinematics  over a certain number of time samples; for instance, $\mat{P} = \left[ \vec{p}(0) \ \vec{p}(1) \ \vec{p}(2)\right]$ is a recording of the generalized position over $3$ time samples.

\subsection{Pseudo-periodicity}%
\label{sec:pseudoperiodicity_baselines}

\begin{figure*}[t]
  \centering
  \subfloat[]{\includegraphics[]{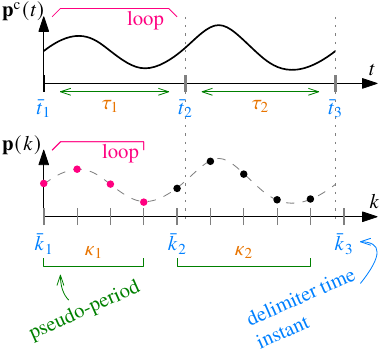}\label{fig:signals_in_time}}
  \hfill 
  \subfloat[]{\includegraphics[]{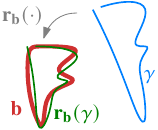}\label{fig:shape_reduction}}
  \hfill
  \subfloat[]{\includegraphics[]{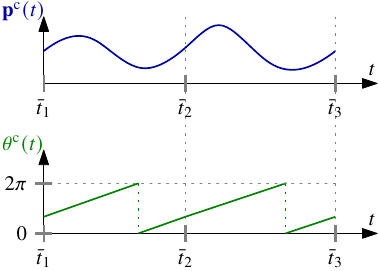}\label{fig:phase_estimation}}
  \caption{(a) Example of a pseudo-periodic signal $\vec{p}^\R{c}$ and its discretization $\vec{p}$ (§ \ref{sec:methods}.\ref{sec:pseudoperiodicity_baselines}).
  (b) Two curves $\vec{\gamma}$ (blue) and $\vec{b}$ (red), and the shape reduction ($\vec{r}_\vec{b}(\vec{\gamma})$) (green) of $\vec{\gamma}$ onto $\vec{b}$ (§ \ref{sec:methods}.\ref{sec:pseudoperiodicity_baselines}).
  (c) Signals involved in the phase estimation problem, Problem \ref{prob:estimate_phase}.}
\end{figure*}

Next, we define the concept of pseudo-periodicity that will allow us to characterize the repetitive nature of different signals for which we wish to perform phase estimation (the Definition is depicted graphically in Figure \ref{fig:signals_in_time}).

\begin{definition}[Pseudo-periodicity]\label{def:pseudoperiodicity}
  Given two sufficiently small constants \emph{($\varepsilon_\R{t}$, $\varepsilon_\R{s}$)}, a continuous-time signal
  $\vec{p}^\R{c}$ is said to be pseudo-periodic over a certain time domain (under distance $\ON{dist}$ and under norm $\norm{\cdot}$) if there exist time instants $\bar{t}_i$ ($i \in \BB{N}_{>0})$ that partition the time domain in intervals $[\bar{t}_i,\bar{t}_{i+1})$---called \emph{pseudo-periods}---of length
  $\tau_i \coloneqq \bar{t}_{i+1} - \bar{t}_i$ such that:
  \begin{enumerate}[label=(\alph*)]
  \item 
\begin{equation}\label{eq:constraint_time}
        \frac{\abs{\tau_{i+1} - \tau_{i}}}{\tau_i} \le \varepsilon_\R{t},
    \end{equation}
    \item

\begin{equation}\label{eq:constraint_value} 
        \frac{\ON{dist}
        \left( 
        \ON{restr}_{[\bar{t}_i, \bar{t}_{i+1})}(\vec{p}^\R{c}),  
        \ON{restr}_{[\bar{t}_i, \bar{t}_{i+1})}(\tilde{\vec{p}}^{\R{c}, \tau_i})
        \right)}
        {\norm{\ON{restr}_{[\bar{t}_i, \bar{t}_{i+1})}(\vec{p}^\R{c})}}
        \le \varepsilon_{\R{s}},
    \end{equation}
where $\tilde{\vec{p}}^{\R{c}, \tau_i}(t) \coloneqq \vec{p}^{\R{c}}(t + \tau_i)$ is a time-shifted version of $\vec{p}^\R{c}$.    
    \end{enumerate}
    The image of $\ON{restr}_{[\bar{t}_i, \bar{t}_{i+1})}(\vec{p}^\R{c})$  is termed as a \emph{loop}.
\end{definition}
In Definition \ref{def:pseudoperiodicity}, \eqref{eq:constraint_time} requires that consecutive pseudo-periods are close in length, while \eqref{eq:constraint_value} enforces that consecutive loops remain similar.
For instance, if the infinity norm were used in \eqref{eq:constraint_value}, letting $p^\R{c}_q$ denote the $q$-th component of $\vec{p}^\R{c}$, we would have
\[
    \frac{\sup_{t \in [\bar{t}_i, \bar{t}_{i+1})} \max_q \abs{p^{\R{c}}_q (t) - p^{\R{c}}_q(t+\tau_i)}}
    {\sup_{t \in [\bar{t}_i, \bar{t}_{i+1})} \max_q \abs{p^{\R{c}}_q(t)}}
    \le \varepsilon_{\R{s}}.
\]
Other meaningful choices for the distance $\ON{dist}$ in \eqref{eq:constraint_value} include those induced by other $L^p$ norms, \emph{dynamic time warping} \cite{senin2008dynamic} or the Fréchet distance \cite{EiterMannila1994}.
Clearly, a signal is \emph{periodic} if and only if it is pseudo-periodic with $\varepsilon_\R{s} = \varepsilon_{\R{t}} = 0$.%
\footnote{Note that Definition \ref{def:pseudoperiodicity} can be also given by comparing not only two consecutive pseudo-periods in \eqref{eq:constraint_time} and (the relative loops) in \eqref{eq:constraint_value}, but rather a number $P$ of them; then, the smaller $P$ is, the more the signal is allowed to drift over time.
Here, we choose $P=2$ for the sake of clarity.}
Next, we extend the definition of pseudo-periodicity to discretized signals (see again Figure \ref{fig:signals_in_time}).

\begin{definition}[Discrete pseudo-periodic signals]
\label{def:pseudoperiodicity_discrete}
    A discrete-time signal $\vec{p}$ is \emph{pseudo-periodic} if it is the discrete-time sampling (cf.~§ \ref{sec:methods}.\ref{sec:notation}; with sampling time $T_\R{s}$) of a continuous time signal $\vec{p}^\R{c}$ that is pseudo-periodic.
    Moreover the \emph{discrete delimiter time instants} are $\bar{k}_i = \arg\min_{k} \abs{\bar{t}_i - k T_\R{s}}$,
    and the \emph{discrete pseudo-periods} are 
    $\C{P}_i \coloneqq \{\bar{k}_i, \dots, \bar{k}_{i+1}-1\}$, with lengths $\kappa_i \coloneqq \bar{k}_{i+1} - \bar{k}_i$.
\end{definition}
A loop of a discrete-time pseudo-periodic signal can be written as a recording: e.g., the $i$-th loop is $[\vec{p}(\bar{k}_i) \ \cdots \ \vec{p}(\bar{k}_{i+1}-1)]$.

\subsection{Baselines}%

Next, we introduce a concept to formalize when a signal is a realization of a specific biological process (e.g., a certain physiotherapy exercise or cardiac activity).
This concept will then allow us to establish meaningful reference points on the signal (e.g., the start of a repetition of the exercise, or the onset of a heartbeat), facilitating phase comparisons across signals that stem from the same underlying process.
Namely, we say that a pseudo-periodic signal \emph{has a baseline} if all loops are similar to that of a specific curve, which is the \emph{baseline}.
We define the concept formally below, and provide a graphical depiction in Figure \ref{fig:shape_reduction} (where a signal loop, $\vec{\gamma}$, is transformed into $\vec{r}_{\vec{b}}(\vec{\gamma})$ to closely resemble the baseline $\vec{b}$).

Let $\C{AT}$ denote the set of all affine transformations (translation, rotation, and scaling) from $\BB{R}^d$ to itself.

\begin{definition}[Shape reduction]
Given two curves $\vec{b}$ and ${\vec{\gamma}}$ with image in $\BB{R}^d$, the \emph{shape reduction} of ${\vec{\gamma}}$ onto $\vec{b}$ is
\[
    \vec{r}_{\vec{b}}(\vec{\gamma})
    = \alpha^\star (\vec{\gamma}),
    \quad
    \text{where }
    \alpha^\star \coloneqq \arg \min_{\vec{\alpha} \in \C{AT}} \ON{dist}_{\R{F}}(\vec{b}, \vec{\alpha}({\vec{\gamma}})).
\]
\end{definition}

\begin{definition}[Baseline]
    A closed invertible curve $\vec{b} : [0, 2 \pi) \rightarrow \BB{R}^d$ is a \emph{baseline} for a continuous-time pseudo-periodic signal $\vec{p}^\R{c}$, with tolerance $\varepsilon_{\R{F}}$, if 
    \[ 
        \ON{dist}_{\R{F}}(\vec{b}, \vec{r}_{\vec{b}}(\ON{restr}_{[\bar{t}_i, \bar{t}_{i+1})}(\vec{p}^\R{c}))) \le \varepsilon_{\R{F}},
        \quad \forall i.
    \]    
\end{definition}

\subsection{Core distance metric}
\label{sec:normalized_distance}

To enable robust comparison across biological signals with different amplitudes and shapes, we introduce the following normalized distance measurement.
Let $\ell \subset \BB{N}_{>0}$ be the length of a signal segment,
$\mat{X} = [\vec{x}(0) \ \cdots \ \vec{x}(\ell-1)]\in \BB{R}^{d \times \ell}$ be a recording, and $\bar{\vec{x}} \in \BB{R}^d$ be a reference point. We define the normalized distance function as:
\begin{equation}\label{eq:normalized_distance}
    \ON{dist}_\R{n}(\mat{X}, h, \bar{\vec{x}}) =  \frac{\norm{\vec{x}(h) - \bar{\vec{x}}}_2}{ \max\limits_{\hbar : \vec{x}(\hbar) \in \mat{X}} \norm{\vec{x}(\hbar) - \bar{\vec{x}}}_2}.
\end{equation}
This normalization scales distances between 0 and 1, making the algorithm robust to amplitude variations that are irrelevant to phase estimation but could otherwise dominate distance calculations. In practice, this means ROPE can track phase consistently whether a patient performs movements with high or low effort, or when sensor sensitivities vary across experimental sessions.

\section{Problem formulation}
\label{sec:problem_formulation}

The mathematical framework established in Section \ref{sec:methods} enables us to address a fundamental challenge in biological signal analysis: specifically, how to assign meaningful phase values to complex, multidimensional oscillatory patterns in real-time.
This section formalizes two related but distinct phase estimation problems that arise naturally in applications.

\subsection{Phase of a signal}

Consider a typical scenario in motor rehabilitation: a stroke patient performs repetitive reaching movements while a therapist monitors their progress. 
Each reach follows a similar three-dimensional trajectory, but natural variability makes the motion pseudo-periodic rather than perfectly periodic.
To track patient's progress and provide feedback at specific phases of the movement, it is useful to abstract away from the exact trajectory of each cycle and instead assess the overall rhythm of the exercise. 

More formally, we seek to estimate  the phase of a signal $\vec{p}^\R{c} : [0, t_{\R{end}}] \rightarrow \BB{R}^d$, which we refer to as the \emph{estimand} signal, where $t_\R{end} \in \BB{R}_{> 0}$ represents the total observation duration.
To ensure this problem is well-posed and computationally tractable, we make the following assumptions about the  signal structure.

\begin{assumption}\label{ass:signals}
    The following conditions hold for the estimand signal $\vec{p}^{\R{c}}$.
    \begin{enumerate}
        \item
        \label{ass:1}
        $\vec{p}^{\R{c}}$ is pseudo-periodic (Definition \ref{def:pseudoperiodicity}; for some tolerance pair $(\varepsilon_\R{t}, \varepsilon_\R{s})$, under the distance induced by the infinity norm and under the infinity norm).
        \item 
        \label{ass:2}
        Within any pseudo-period there are no self-intersections; i.e., $\vec{m}^\R{c}(t_1) \ne \vec{m}^\R{c}(t_2)$ for any different $t_1$, $t_2$ in the pseudo-period).
        \item
        \label{ass:3}
        All pseudo-periods lengths are upper-bounded by a known constant $\tau_{\R{max}}\in \BB{R}_{>0}$; i.e., $\tau_i \le \tau_{\R{max}}, \forall i$.
        \item
        \label{ass:4}
        Discrete-time kinematic measurements $\vec{m}(k)$ are available.
    \end{enumerate}  
\end{assumption}

These assumptions reflect realistic constraints for biological signals. 
Point \ref{ass:1} captures the natural variability in biological rhythms, while point \ref{ass:2} ensures that each point in a movement cycle can be uniquely identified, a reasonable requirement for most biological motions; nonetheless, for signals that do not satisfy to this assumption (e.g., for displaying stationary phases) we relax the assumption in Appendix \ref{sec:relaxation_ass_intersections}. 
Point \ref{ass:3} provides the computational constraint needed for real-time implementation, and point \ref{ass:4} reflects the need for measurements of the oscillatory signal of interest.

With these foundations, we can formulate our primary phase estimation problem (see Figure \ref{fig:phase_estimation}).

\begin{problem}[Phase estimation]\label{prob:estimate_phase}
    Design a function $\theta^\R{c} : [0, t_{\R{end}}] \rightarrow \BB{S}$ such that within each pseudo-period $[\bar{t}_i, \bar{t}_{i+1})$, the phase function $\theta^\R{c}$ (i) is bijective, (ii) has constant derivative, and (iii) contains at most one discontinuity.
\end{problem}

The sawtooth phase function provides a natural solution to  Problem \ref{prob:estimate_phase} (see Figure \ref{fig:phase_estimation}. Within each detected cycle, phase increases linearly from an initial offset value to that offset plus $2\pi$:
\begin{equation}\label{eq:form_phase_function}
    \quad
    \theta^\R{c}(t) = \left[ \frac{2\pi (t - \bar{t}_i)}{\bar{t}_{i+1}-\bar{t}_i} + \theta_{\R{offset}} \right] \bmod 2 \pi,
\end{equation}
where $\bar{t}_i$ and $\bar{t}_{i+1}$ are consecutive delimiter time instants that mark cycle boundaries (cf.~Definition \ref{def:pseudoperiodicity}). 
To compute $ \theta^\R{c}(t)$ from \eqref{eq:form_phase_function},  we must estimate the delimiter time instants and specify the phase offset
$\theta_\R{offset} \in \BB{S}$.
While setting $\theta_\R{offset}=0$ works for analyzing individual signals, different offset values become crucial when comparing multiple signals, as we explore next in Section \ref{sec:problem_formulation}.\ref{sec:multiple_signals}. 
The real-time implementation challenge lies in maintaining piecewise-linear phase derivatives, which requires predicting when the next cycle boundary $\bar{t}_{i+1}$ will occur before it happens. This is a non-trivial task given the natural variability in biological cycle durations.

\subsection{Phase alignment for multiple signals}
\label{sec:multiple_signals}

\begin{figure*}
    \centering
    \subfloat[]{\includegraphics[]{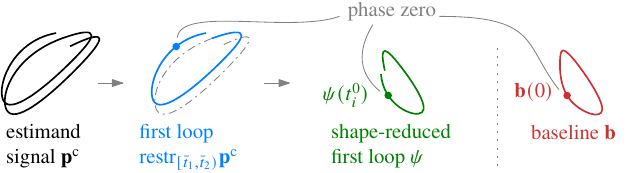}\label{fig:zero_phase}}
    \hfill
    \subfloat[]{\includegraphics[]{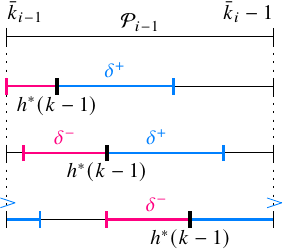}\label{fig:search_interval}} 
    \caption{%
    (a) Points associated to phase zero, on the baseline $\vec{b}$ and on the estimand signal $\vec{p}^\R{c}$ (see § \ref{sec:problem_formulation}.\ref{sec:multiple_signals}).
    (b) Search domains used in \eqref{eq:minimization_search} and \eqref{eq:smaller_search_domain}.    
    The first horizontal line reports the full search domain in \eqref{eq:minimization_search}, the second and third lines depict the first case in
    \eqref{eq:smaller_search_domain}, whereas the last line depicts the second case in \eqref{eq:smaller_search_domain}.}
\end{figure*}

Many biological applications require comparing phase relationships between multiple oscillatory systems. 
Consider multi-person coordination studies or cardiac-respiratory coupling analysis, where we need to determine whether breathing and heart rate are synchronized.
Simply applying Problem \ref{prob:estimate_phase} to each signal independently would assign phase zero arbitrarily, making comparisons meaningless. 
Instead, a consistent phase reference is needed to align corresponding physiological events.
This challenge arises whenever multiple pseudo-periodic signals share similar underlying patterns, formalized as having a common baseline $\vec{b}$. 
In such cases, we aim to determine whether the signals exhibit phase synchronization (coordinated timing) or frequency synchronization (matched cycle rates).
%
%
To assess frequency synchronization, it suffices to solve Problem \ref{prob:estimate_phase} for each signal independently and compare the derivatives of their phase functions.
Phase synchronization, however, requires a stricter condition: for all signals, phase zero must be assigned to the point that, after shape reduction, lies closest to the baseline's zero-phase point (see Figure \ref{fig:zero_phase}).
This alignment ensures that phase comparisons reflect true coordination of biological events.

Formally, we define the transformed first loop $\vec{\psi} \coloneqq \vec{r}_\vec{b}(\ON{restr}_{[\bar{t}_1, \bar{t}_2)}(\vec{p}^\R{c}))$, which represents the estimand signal $\vec{p}^\R{c}$ within its first pseudo-period, geometrically transformed via an affine mapping to match, as closely as possible, the shape of the baseline $\vec{b}$.
We then identify $\vec{q}^\star \coloneqq \arg \min_{\vec{q} \in \ON{image}(\vec{\psi})} \norm{\vec{q} - \vec{b}(0)}$ as the point on the transformed loop closest to the baseline's phase-zero reference point (cf.~Figure \ref{fig:zero_phase}).
We also denote $t^0$ as the time instant within $[\bar{t}_1, \bar{t}_2)$ corresponding to phase zero, that is 
$t^0 \coloneqq (\ON{restr}_{[\bar{t}_1, \bar{t}_2)}(\theta^\R{c}))^{-1}(0)$.
These definitions enable us to formulate the following phase alignment problem:

\begin{problem}\label{prob:zero_phase}
    Solve Problem \ref{prob:estimate_phase} with the additional constraint that $\theta^\R{c}$ satisfies
    $
        \vec{\psi}(t^0) = \vec{q}^\star
    $.    
\end{problem}
Note that Problem \ref{prob:zero_phase} can be solved by appropriately selecting $\theta_\R{offset}$  in \eqref{eq:form_phase_function}, ensuring that phase zero aligns with the designated reference point in each cycle. 

To ensure that Problem \ref{prob:zero_phase} has a well-defined solution, we require the following assumption.
\begin{assumption}\label{ass:baseline}~
  \begin{enumerate}
    \item The baseline $\vec{b}$ is known.
    \item 
    Either $\vec{b}$ and $\vec{p}^\R{c}$ are expressed in the same frame of reference, or the coordinate transformation between the signal and baseline reference frames is known.
  \end{enumerate}
\end{assumption}
Assumption \ref{ass:baseline} is required to allow computation of $\vec{r}_\vec{b}$. 
This assumption is reasonable for many biological applications where reference patterns can be established from healthy subjects or theoretical models, and where experimental protocols ensure consistent coordinate systems across measurements.

\section{Algorithm description} \label{Sec:algorithm}

We are now ready to present ROPE (Recursive Online Phase Estimation), our solution to the phase estimation problem formalized in Section \ref{sec:problem_formulation}.
The algorithm’s design is guided by a key biological insight: current physiological states tend to resemble 
those of the same type from recent cycles.
For example, a heartbeat in progress is most similar to the corresponding moment in the previous heartbeat, rather than to arbitrary past points.
ROPE leverages this by storing recently completed cycles and using pattern matching to locate the current state within the cycle template.
This approach accommodates the variability inherent in biological systems while providing the real-time performance essential for clinical and research applications.

\subsection{Algorithm overview and operating modes}

\begin{figure*}[t]
   \centering
   \subfloat[]{\includegraphics[scale=0.96]{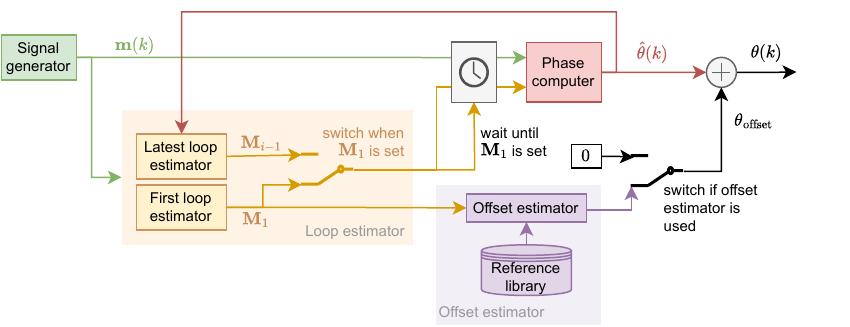}
   \label{fig:scheme_algorithm}}%
   \hspace{1.5cm} 
   \subfloat[]{\includegraphics[scale=0.96]{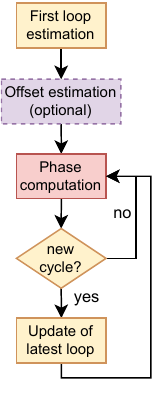}
   \label{fig:flowchart}}
   \caption{(a) Block scheme of the phase estimator for multidimensional periodic motion.
   (b) Flowchart of the phase estimation algorithm.}
\end{figure*}

To enable implementation on digital computers, ROPE operates on a discretized signal  $\vec{m} \in [0, k_{\R{end}}] \to \BB{R}^d$ ($k_\R{end} \in \BB{N}_{> 0}$) and outputs the discrete-time phase function $\theta : [0, k_{\R{end}}] \to \BB{S}$.
Figure \ref{fig:scheme_algorithm} illustrates the algorithm's modular architecture, designed for flexibility across diverse biological applications, while Figure \ref{fig:flowchart} depicts a flowchart representation of the steps involved in the algorithm.

ROPE supports two complementary operating modes 
which we term as \emph{untethered} and \emph{tethered}.
The untethered mode solves Problem \ref{prob:estimate_phase} by estimating phase without requiring an absolute reference; it is suited for applications involving individual subjects or systems where relative phase tracking is sufficient, such as monitoring a patient's movement consistency during rehabilitation.
Conversely, the tethered mode solves Problem \ref{prob:zero_phase}, ensuring consistent phase alignment across multiple signals; it uses the additional ``offset estimator'' block and requires Assumption \ref{ass:baseline}.
This mode is essential when comparing signals or interpreting phase relative to physiological landmarks, such as aligning movements in coordination studies or synchronizing cardiac and respiratory phases.
The modular design allows end-users to select the mode best suited to their objectives, with the algorithm automatically adapting its computational strategy.


\subsection{Algorithm components}
\label{sec:algorithm_components}

 Next, we describe each of ROPE components.
 Their input-output relationships are illustrated in the block diagram in Figure \ref{fig:scheme_algorithm}, whereas a flowchart is reported in Figure \ref{fig:flowchart}.

\subsubsection{Detection of the first cycle}

Accurate phase estimation fundamentally depends on correctly identifying when one biological cycle ends and the next begins. ROPE addresses this challenge through the ``loop estimator'' (see Figure \ref{fig:scheme_algorithm}).
When analyzing a new biological signal, ROPE must first identify a complete initial cycle without prior information (detection of subsequent cycles follows a different principle, as explained in Section \ref{Sec:algorithm}.\ref{sec:algorithm_components}.\ref{sec:detection_subsequent_cycles}).
The ``first loop estimator'' addresses this problem by analyzing a sufficiently long initial recording
\[
  \mat{M}_0 = \begin{bmatrix}
    \vec{m}(0) & \cdots & \vec{m}(\kappa_0-1)
  \end{bmatrix} \in \BB{R}^{d \times \kappa_0}
\]
of (integer) length $\kappa_0 > 2 \tau_{\R{max}} / T_{\R{s}}$, ensuring the recording includes at least two pseudo-periods.
The algorithm then searches for the cycle length that maximizes similarity between consecutive signal segments, effectively identifying the fundamental repetition period in the biological data.
The computation accounts for the multidimensional nature of biological signals through normalized cross-correlation analysis.
As we take $\bar{k}_1 = 0$, to identify the first pseudo-period, we need to determine the value $\bar{k}_2$ that maximizes similarity between the first $n$ samples and the subsequent $n$ samples across all signal dimensions. 

To perform the computation, we let $c_n$ denote the correlation between the first $n$ columns of $\mat{M}_0$ and its subsequent $n$ columns, and we pick as $\bar{k}_2$ the value of $n$ that maximizes this correlation; see Figure \ref {fig:first_loop_estimation}.
Formally, let $\mat{M}_{0\mid i:j}$ denote the submatrix obtained from $\mat{M}_0$  by picking columns from the $i$-th to the $j$-th (both included).
Let $\vec{\mu}(\mat{A})$ (resp.\ $\vec{\sigma}(\mat{A})$) return a column vector where the $i$-th element is the mean (resp.\ standard deviation) of the $i$-th row of a matrix $\mat{A}$.
Additionally, we let 
$\bar{\vec{\mu}}_{2n} \coloneqq \vec{\mu}(\mat{M}_{0\mid 1:2n}) \in \BB{R}^d$,
and 
$\bar{\mat{\Sigma}}_{2n} \coloneqq \frac{1}{n} \ON{diag} \left( \vec{\sigma}(\mat{M}_{0\mid 1:2n})^{\circ -2}\right) \in \BB{R}^{d \times d}$.
Then,
\[
  c_n \coloneqq \ON{tr}(
  [\mat{M}_{0\mid 1:n} - \bar{\vec{\mu}}_{2n} \vec{1}_n\T]
  [\mat{M}_{0\mid n+1:2n} - \bar{\vec{\mu}}_{2n} \vec{1}_n\T]\T \mat{\Sigma}).
\]
Hence, we take
\[
  \bar{k}_2 = \arg \max_{n \in \left( 1, \dots, \left\lfloor \frac{\kappa_0}{2} \right\rfloor \right)} c_n.
\]

\subsubsection{Real-time phase computation}

\begin{figure*}
    \centering
    \subfloat[]{\includegraphics[]{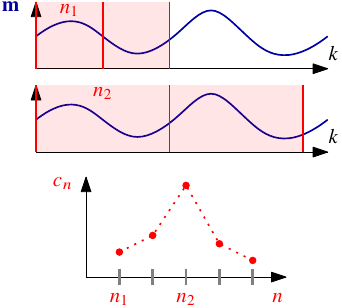}\label{fig:first_loop_estimation}}
    \hfill
    \subfloat[]{\includegraphics[]{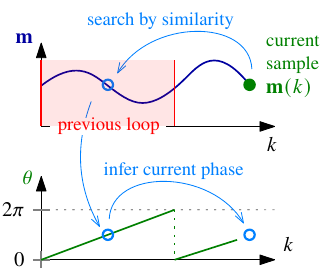}\label{fig:search_previous_loop}}
    \hfill
    \subfloat[]{\includegraphics[]{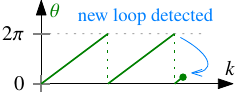}\label{fig:new_loop}}
    \caption{%
    Representation of key steps involved in the ROPE algorithm, as explained in Section \ref{Sec:algorithm}.
    (a) Process performed by the ``loop estimator'' block, in the estimation of the first loop.
    (b) Process performed by the ``Phase computer '' block.
    (c) Process performed by the ``loop estimator'' block, in the estimation of the second and later loops.}
\end{figure*}

Once the initial cycle is identified by the ``first loop estimator,'' the ``phase computer'' can start estimating the current phase by finding the historical moment that best matches the present signal state, as portrayed graphically in Figure \ref {fig:search_previous_loop}. 
In the $i$-th pseudo-period $\C{P}_{i}$ (see Definition \ref{def:pseudoperiodicity_discrete}), at each time instant $k \in \C{P}_{i}$, the ``phase computer'' block takes two key inputs: the current multidimensional kinematic signal $\vec{m}(k)$ (cf.\ § \ref{sec:methods}.\ref{sec:notation}) and the latest available loop:
\[
    \vec{M}_{i-1} = \begin{bmatrix}
    \vec{m}(\bar{k}_{i-1}) & \cdots & \vec{m}(\bar{k}_{i}-1)
  \end{bmatrix},
  \]
where the delimiter time instants $\bar{k}_{i-1}$, $\bar{k}_{i}$ are provided by the ``period estimator''  (described below). 
The algorithm then searches for the historical point that most closely resembles the current state by minimizing a combined distance metric (cf.~§ \ref{sec:methods}.\ref{sec:normalized_distance}) that considers both position and velocity information:
\begin{equation}\label{eq:minimization_search}
  h^*(k)  \coloneqq  \arg \min_{h \in \C{P}_{i-1}} \left[ 
  \ON{dist}_\R{n} (\mat{P}_{i-1}, h, \vec{p}(k)) + 
  \ON{dist}_\R{n} (\mat{V}_{i-1},h,\vec{v}(k)) \right],
\end{equation}
where positions ($\mat{P}_{i-1}$, $\vec{p}(k)$) and velocities ($\mat{V}_{i-1}$, $\vec{v}(k)$) are extracted from the  kinematics data ($\mat{M}_{i-1}$, $\vec{m}(k)$).
Notably, searching for a match to the current signal state in the previous loop $\C{P}_{i-1}$---rather than in a fixed reference---allows the algorithm to remain flexible and adapt to drifting signals.
Once the best matching point $h^*(k)$ is identified, ROPE computes the phase using the sawtooth function established in  \eqref{eq:form_phase_function} by setting:%
\footnote{Crucially, differently from \eqref{eq:form_phase_function}, we use the delimiter time instants of the ($i-1$)-th pseudo-period (i.e., $\bar{k}_{i-1}$, $\bar{k}_{i}$), rather than those of the $i$-th one (i.e., $\bar{k}_{i}$, $\bar{k}_{i+1}$), to allow real time computation, as $\bar{k}_{i+1}$ is unknown before the $i$-th pseudo-period is completed.}
\begin{equation}\label{eq:phase_value}
    \theta^\R{c}(k) = \left[ \frac{2\pi (h^*(k) - \bar{k}_{i-1})}{\bar{k}_{i}-\bar{k}_{i-1}} + \theta_{\R{offset}} \right] \bmod 2 \pi.
\end{equation}
In the untethered mode, $\theta_\R{offset}$ is set to $0$; in the tethered mode, the ``offset estimator'' (described in § \ref{Sec:algorithm}.\ref{sec:algorithm_components}.\ref{sec:offset_estimator}) sets this parameter to ensure consistent phase alignment across multiple signals.

Without specific assumptions on $\mat{P}_{i-1}$ and $\mat{V}_{i-1}$, the computation time required to solve \eqref{eq:minimization_search} is proportional to $\abs{\C{P}_{i-1}} = \bar{k}_{i}-\bar{k}_{i-1}$.
In the Appendix \ref{sec:accelerated_minimization_search}, we present an optional method to reduce this computation time.

\subsubsection{Detection of the second and later cycles}
\label{sec:detection_subsequent_cycles}

After the initial cycle is established and phase estimates are being provided by the ``phase computer'', ROPE switches to a more efficient transition-based detection strategy. 
The ``latest loop estimator'' identifies new cycle boundaries by monitoring phase transitions. Specifically, to determine the $i$-th pseudo-period with $i > 1$, we need to find $\bar{k}_{i+1}$.
This is estimated as the discrete time instant $k^*$ such that, according to \eqref{eq:phase_value}, $\theta({k^*}) - \theta({k^*-1}) < - \pi$, indicating a phase jump from $2\pi$ to $0$ (see Figure \ref{fig:new_loop}).
This transition-based approach provides the computational efficiency necessary for real-time operation while maintaining the accuracy established during the initial cycle detection.

\subsubsection{Phase alignment for multiple signals}
\label{sec:offset_estimator}
When operating in tethered mode, ROPE must ensure that phase zero corresponds to the same physiological event across all analyzed signals.
The ``offset estimator'' addresses this challenge by geometrically aligning the first detected cycle with a reference baseline pattern $\vec{b}$ (cf.~Figure \ref{fig:zero_phase}), to compensate for  differences in coordinate systems, signal scaling, and spatial orientation.
This alignment process involves a sequence of rotation, translation, and scaling, performed to homogenize the observed cycle to the baseline; then the algorithm identifies the phase-zero point on the baseline and computes the appropriate $\theta_\R{offset}$ value that ensures consistent phase assignment across the various biological signals. 

Namely, the ``offset estimator'' block performs its computation once, as soon as $\bar{k}_2$ is estimated and the first loop $\mat{M}_1$ is available (see Figure \ref{fig:flowchart}). 
The block takes as input the first (position) loop, i.e.,
\[
  \mat{P}_1 = \begin{bmatrix}
    \vec{p}(0) & \cdots & \vec{p}(\bar{k}_2 - 1)
  \end{bmatrix}
  \in \BB{R}^{d \times \kappa_1},
\]
the reference frame $\C{R}_{\R{e}} = \{ \vec{o}_\R{e}, \vec{u}_{\R{e},1}, \ldots, \vec{u}_{\R{e},d}\}$ in which it is expressed, a recording of the baseline position, i.e.,
\[
  \mat{P}_\R{b} = \begin{bmatrix}
    \vec{p}_\R{b}(0) & \cdots & \vec{p}_\R{b}(\kappa_\R{b}-1)
  \end{bmatrix}
  \in \BB{R}^{d \times \kappa_\R{b}},
\]
and the reference frame $\C{R}_{\R{b}} = \{ \vec{o}_\R{b}, \vec{u}_{\R{b},1}, \ldots, \vec{u}_{\R{b},d}\}$ in which it is expressed.
The block outputs the phase offset $\theta_\R{offset}$, which is computed as follows.

The rotation matrix that transforms 
$\C{R}_\R{e}$ into $\C{R}_{\R{b}}$ is 
\[
  \mat{R} = 
  \begin{bmatrix}
      \vec{u}_{\R{b},1} & \cdots & \vec{u}_{\R{b},d}
  \end{bmatrix}
  \begin{bmatrix}
      \vec{u}_{\R{e},1} & \cdots & \vec{u}_{\R{e},d}
  \end{bmatrix}^{-1}.
\]
First, we rotate $\mat{P}_1$, obtaining $\mat{P}_1^{\R{r}} = \mat{P}_1 \mat{R}\T$.
Second, we denote by $\vec{c} \in \BB{R}^d$ the centroid of $\mat{P}_1^{\R{r}}$ (each element of $\vec{c}$ is the average of the corresponding row in $\mat{P}_1^{\R{r}}$), and compute the translated recording
$\mat{P}_1^{\R{r}, \R{t}} = \mat{P}^{\R{r}}-\vec{c}$.
Third, we perform a scaling as follows.
Let 
$\tilde{\vec{\sigma}} \coloneqq 
\vec{\sigma} (\mat{P}_1^{\R{r}, \R{t}})
\circledslash
\vec{\sigma} (\mat{B})$,
the scaled recording is then computed as
$\mat{P}_1^{\R{r}, \R{t}, \R{s}} = \mat{P}_1^{\R{r}, \R{t}} \odot
\left( \tilde{\vec{\sigma}} \vec{1}_{\kappa_1}\T\right)$.%
\footnote{Referencing Section \ref{sec:problem_formulation}.\ref{sec:multiple_signals}, $\mat{P}_1$ corresponds to (the discretization of) $\ON{restr}_{[\bar{t}_1, \bar{t}_2)}(\vec{p}^\R{c})$, while
$\mat{P}_1^{\R{r}, \R{t}, \R{s}}$ corresponds to (the discretization of)
$\vec{\psi}$.}
Finally, we let $\vec{p}^{\R{r}, \R{t}, \R{s}}(0) \in \BB{R}^{d}$ denote the first column of $\mat{P}_1^{\R{r}, \R{t}, \R{s}}$ and compute $\vec{v}^{\R{r}, \R{t}, \R{s}}(0) \in \BB{R}^{d}$ as its numerical derivative. 
We compute $\vec{v}_\R{b}$ as the numerical derivative of $\vec{p}_{\R{b}}$, and define the baseline velocity loop as
\[
  \vec{V}_{\R{b}} \coloneqq \begin{bmatrix}
    \vec{v}_\R{b}(0) & \cdots & \vec{v}_\R{b}(\kappa_\R{b}-1)
  \end{bmatrix}
  \in \BB{R}^{d \times \kappa_\R{b}}.
\]
Then, similarly to \eqref{eq:minimization_search}, we compute
\begin{equation}\label{eq:minimization_search_offset}
  h^*_0  \coloneqq  \arg \min_{  h \in \{0, \dots, \kappa_{\R{b}}-1\} } \left\{ \mathrm{dist}(\mat{P}_{\R{b}},h,\vec{p}^{\R{r}, \R{t}, \R{s}}(0))+\mathrm{dist}(\mat{V}_{\R{b}},h,\vec{v}^{\R{r}, \R{t}, \R{s}}(0))) \right\},
\end{equation}
and then we set
\begin{equation}
  \theta_\R{offset} = \frac{2 \pi h^*_0}{\kappa_{\R{b}}} \bmod 2\pi.
\end{equation}

\section{Validation strategy and experimental design}
\label{sec:validation_setup}

Validating a phase estimation algorithm for biological applications requires demonstrating robust performance across the full spectrum of signal complexity encountered in real-world scenarios, including noise, varying conditions, and structured geometries.
To this end, as detailed in Section \ref{sec:validation_setup}.\ref{Sec:Dataset}, we consider three complementary signal categories, progressing from well-characterized synthetic systems (chaotic oscillators) to increasingly challenging biological signals (human movement data and cardiovascular signals).

In all cases, we compared the phase estimate obtained by our ROPE algorithm against a benchmark phase, obtained via careful human inspection, and against the estimates obtained via two different widely used algorithms, termed here as PCA-H and PCA-T, illustrated in Section \ref{sec:validation_setup}.\ref{Sec:Reference algorithm}.


\subsection{Dataset description and experimental protocols} \label{Sec:Dataset}


\subsubsection{Chaotic dynamics: R\"ossler oscillator}  
The first validation scenario involves estimating the phase of a R\"ossler oscillator \cite{Rssler1976}, a classical chaotic nonlinear system exhibiting deterministic chaos, which can arise in biological systems experiencing pathological dynamics or operating near critical transitions, and whose state trajectories can be challenging to map onto a one-dimensional phase signal.
Moreover, the controlled nature of this setup allows us to generate arbitrarily long data sequences for thorough statistical analysis, and systematically vary chaos parameters to test algorithm robustness across different complexity regimes.

The system dynamics follow:
\begin{equation}\label{eq:rossler}
 \dot{x} = -y - z,  
 \quad
 \dot{y} = x + a y, 
 \quad
 \dot{z} = b + z (x - c).
\end{equation}
We generated $5$ distinct chaotic trajectories using different parameter combinations chosen to produce a variety of pseudo-periodic three-dimensional patterns.\footnote{
For simulation 1, we used: $a = 0.2$, $b = 0.2$, $c = 3.7$;
for simulation 2: $a = 0.1 $, $b =  0.1 $, $c = 9.0$;
for simulation 3: $a = 0.15$, $b =  0.25$, $c = 5.5$;
for simulation 4: $a = 0.18$, $b =  0.22$, $c = 4.9$;
for simulation 5: $a = 0.12$, $b =  0.28$, $c = 9.15$.
}
One loop of the fifth recording is portrayed in Figure \ref{fig:shapes}.


\subsubsection{Human movement: motor control and coordination patterns}  

The second scenario focuses on repetitive hand motion performed by human subjects, which is clinically relevant in applications such as rehabilitation and sports science. 
The high cycle-to-cycle variability and intricate shapes typically featured by human motion data pose significant challenges for phase estimation, even causing traditional methods to fail.

\begin{figure*}[t]
  \centering
  \subfloat[]{%
    \includegraphics[max width=\textwidth]{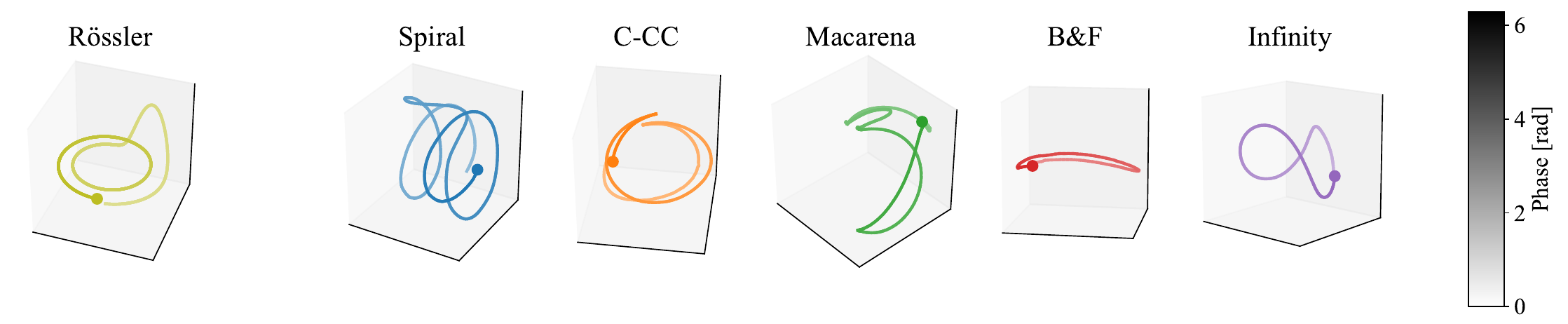}%
    \label{fig:shapes}
  }
  \hfill
  \subfloat[]{%
    \includegraphics[scale=0.50]{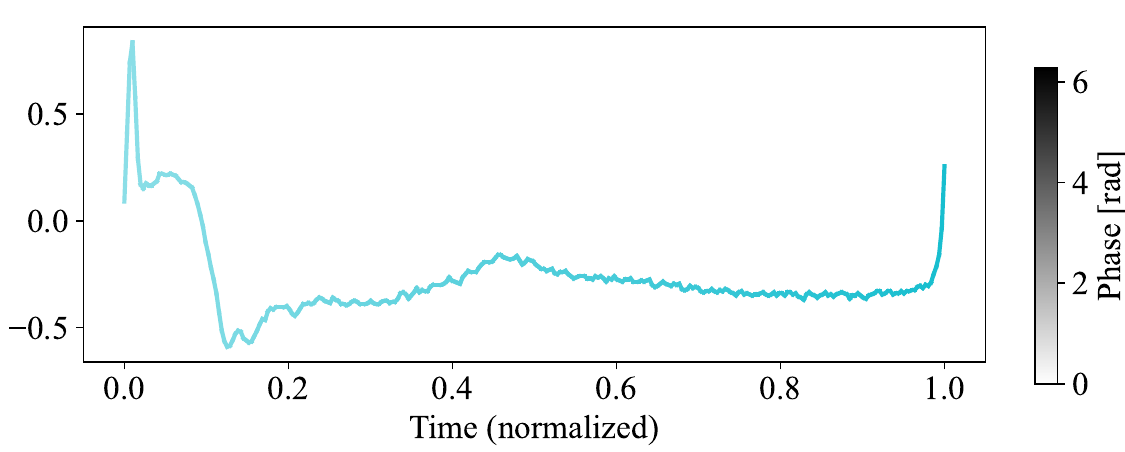}%
    \label{fig:ECG_shapes}
  }
  \caption{(a) Representative loops of the validation signals used in this study: Rössler (chaotic dynamics), Spiral, Clockwise-Counterclockwise (C-CC), Macarena, Back \& Forth (B\&F), and Infinity.
  (b) A representative loop of the ECG signals.}
  \label{fig:multifig_loops}
\end{figure*}

\begin{figure*}[t]
  \centering
  \subfloat[]{%
    \includegraphics[scale=0.195]{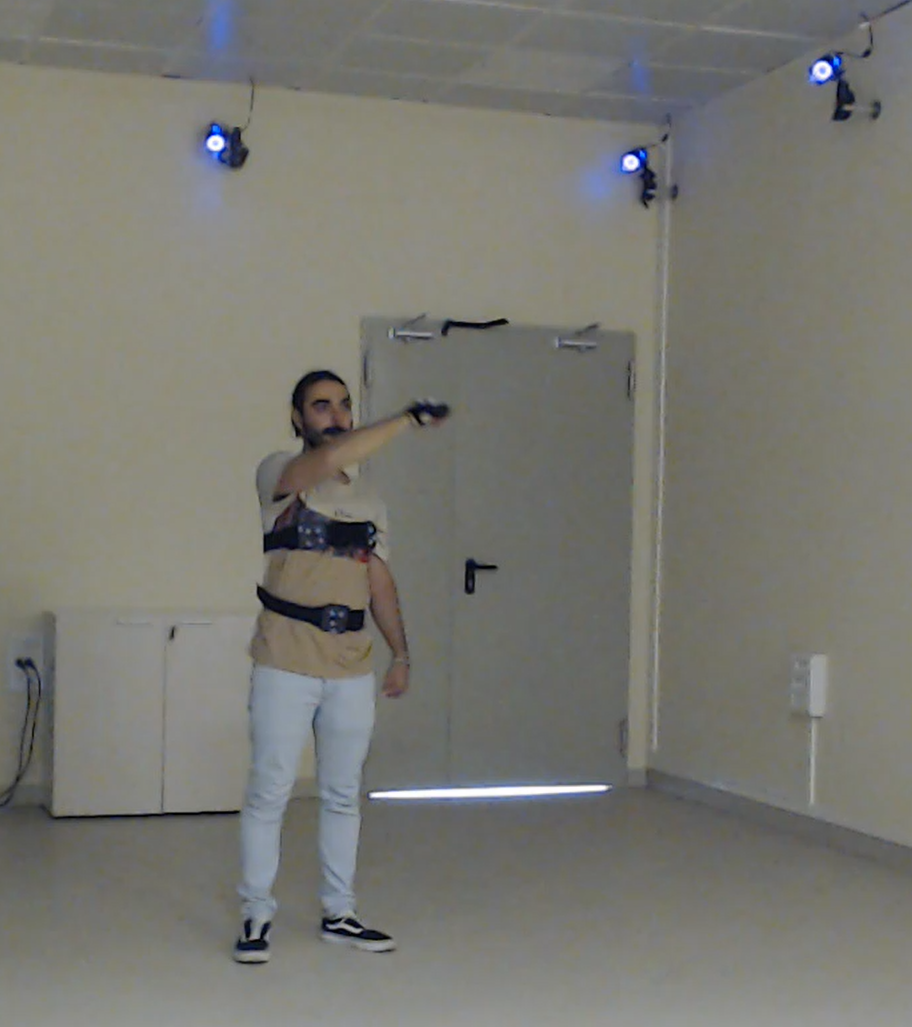}%
    \label{fig:vicon_antonio}
  }
  \hspace{1.5cm} 
  \subfloat[]{%
    \includegraphics[scale=0.2]{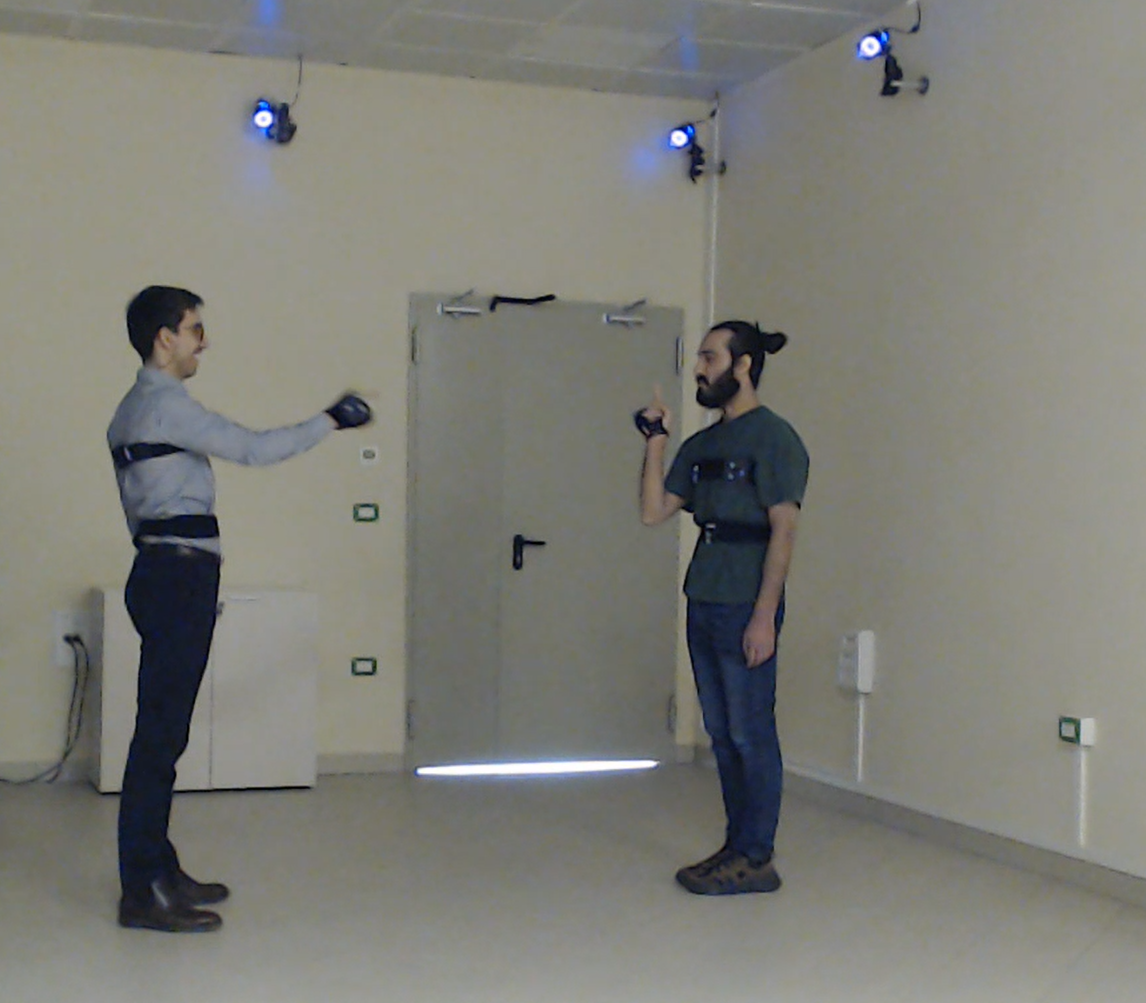}%
    \label{fig:vicon_foto}
  }
  \caption{Photographs of the motion capture setup used to acquire data of human motion signals.
  (a) A participant performing the Spiral movement.
  (b) Two participants performing the Back \& Forth (B\&F) movement while attempting to maintain an anti-phase synchronization.}
  \label{fig:multifig_vicon}
\end{figure*}

This dataset we collected consists of recorded motion trajectories of human subjects, acquired using a Vicon motion-capture system equipped with 10 cameras, ensuring high spatial and temporal accuracy.
$5$ types of motion were performed, spanning different levels of geometric complexity; they are termed as
``Spiral'',
``Clockwise-Counterclockwise'',
``Macarena'',
``Back and Forth'',
``Infinity'', and are illustrated in Figure~\ref{fig:shapes}.
For each motion type, we recorded 9 trials, each lasting 30 seconds, and sampled at 100 Hz.
These motion signals are three-dimensional, as they represent physical trajectories in 3D space. 
Photos representing the motion capture setup are reported in Figure \ref{fig:multifig_vicon}.

\subsubsection{Cardiovascular signals: clinical ECG analysis}  

As the third validation scenario, we consider electrocardiogram (ECG) data, a critical class of physiological signals in clinical practice. 
Their frequently high noise levels make phase estimation difficult even with standard offline techniques, thus serving as a valuable test case.

We used real ECG signals from the MIT-BIH Arrhythmia Database, available on PhysioNet \cite{PhysioNet}. 
The database contains 48 fully annotated two-lead ECG recordings, each lasting approximately 30 minutes, sampled at 360 Hz, and collected from 47 subjects.
We selected 7 pseudo-periodic segments of 30 seconds each from the available signals.
In this case, the signals are one-dimensional.
A representative example is displayed in Figure~\ref{fig:ECG_shapes}.

\paragraph{Verification of Assumption \ref{ass:signals}} 

We verified Assumption \ref{ass:signals} for all the signals in the dataset.
Namely, we computed numerically that Assumption \ref{ass:signals}.\ref{ass:1} holds (under the distance induced by the infinity norm and under the infinity norm), with 
\((\varepsilon_\R{t}, \varepsilon_\R{s}) \in [0.001, 0.124] \times [0.089, 0.582]\) for the Rössler trajectories, 
\((\varepsilon_\R{t}, \varepsilon_\R{s}) \in [0.029,1.063] \times [0.069,0.401]\) for the human motion signals,
and \((\varepsilon_\R{t}, \varepsilon_\R{s}) \in [0.053, 0.091] \times [1.524, 1.826]\) for the ECG signals.
The Rössler trajectories and human motion signals are designed so that Assumption \ref{ass:signals}.\ref{ass:2} holds; however this cannot be ensured for the ECG signals, which are expected to contain portions where the signal remains approximately constant (see Figure \ref{fig:ECG_shapes}).
Hence, when estimating the phase of ECG signals, we modify the optimization problem in \eqref{eq:minimization_search} to remove the need for this assumption, as detailed in Appendix~\ref{sec:relaxation_ass_intersections}.%
\footnote{The downside of this approach is that the algorithm becomes less flexible to changes in the lengths $\tau_i$ of consecutive pseudo-periods.}
%
%
Assumption \ref{ass:signals}.\ref{ass:3} is also satisfied, with 
$\tau_{\max} = 12.27\,\text{s}$ for the Rössler trajectories,
$\tau_{\max} = \{4.60, 3.46, 7.70, 1.32, 3.28\}\,\text{s}$ for the human motion signals (Spiral, Clockwise-Counterclockwise, Macarena, Back \& Forth, and Infinity, respectively),
and $\tau_{\max} = 0.88\,\text{s}$ for the ECG signals.

\subsection{Algorithm validation framework}%
\label{Sec:Reference algorithm}

To rigorously validate ROPE, we compare it against both ground-truth references and established methods, described below. 

\paragraph{Benchmark}
\label{par::banchmark}
To obtain an accurate benchmark value for the phase, we carefully inspected each signal in the dataset and chose delimiter time instants $\bar{k}_i$ that yield low values of $\varepsilon_\R{t}$ and $\varepsilon_\R{s}$ (cf.~Definition \ref{def:pseudoperiodicity}).
Then, we built the benchmark phase signals according to a time-discretization of \eqref{eq:form_phase_function}.

\subsubsection{Comparative phase estimation methods}

We selected two widely-used comparison methods that represent different algorithmic philosophies and are commonly employed in biological signal processing: we refer to them here as PCA-H and PCA-T.

PCA-H [Principal component analysis (PCA) with Hilbert transform] represents the traditional approach to multidimensional phase estimation.
Given a signal $p(t)$, this method first reduces signal dimensionality through PCA, then applies the Hilbert transform to estimate phase as $\theta_{\text{PCA-H}}(t) = \arctan\left(\C{H}(p(t))/p(t)\right)$, where $\C{H}$ denotes such transform \cite{Pikovsky2001}.
While computationally elegant, this approach has two key limitations: the Hilbert transform requires complete signal records, preventing real-time operation, and PCA-based dimensionality reduction may discard critical information, leading to inaccurate phase estimation, as shown in Section \ref{Sec::results}.

On the other hand, PCA-T (PCA and inverse tangent) is used to estimate online the phase of a multidimensional signal $\vec{p}$ by computing, or updating, every $T_\R{update}$ seconds, the principal component (in the sense of the PCA) of the signal, exploiting only its most recent $T_\R{memory}$ seconds.
Then, one obtains a scalar signal $p_{\R{pc}}$ by projecting, at each time step, $\vec{p}$ onto its principal component.
Finally, the phase is estimated as
$\theta_{\text{PCA-T}}(t) = \arctan\left(\dot{x}_{\R{pc}}(t)/(-x_{\R{pc}}(t))\right)$, where $\dot{x}_{\R{pc}}$ is computed numerically \cite{Mrtl2012,grotta2024learningbaseda}.
Both $T_\R{update}$ and $T_\R{memory}$ are tunable parameters.%
\footnote{%
Small values of $T_\R{update}$ imply higher responsiveness and adaptability to chancing conditions, but are computationally more intensive; $T_\R{memory}$ needs to be large enough to capture whole pseudo-periods, but a too-large value compromise adaptability when signals change in time.}
Although using the inverse tangent instead of the Hilbert transform enables online phase estimation, it can deteriorate performance due to the more simplistic nature of the approach (as shown in Section \ref{Sec::results}.\ref{sec:results_untethered} below).


\section{Results and performance analysis} \label{Sec::results}
Our comprehensive validation reveals that ROPE achieves superior phase estimation performance across all tested biological signal types, with particularly pronounced advantages in the challenging real-world conditions that characterize clinical and research applications. 
The results demonstrate not merely statistical superiority, but practically meaningful improvements that translate directly to enhanced biological insights and clinical utility.

\paragraph{Estimation error}

To assess quantitatively the phase estimation performance, we employ the \emph{estimation error}, defined at time $k$ for a certain algorithm as
$
    \varepsilon(k) = \abs{ \angle e^{i(\theta_{\R{bmk}}(k) - \theta_{\R{est}}(k))} },
$
where $\theta_{\R{bmk}}$ is the manually verified benchmark phase and $\theta_{\R{est}}$ is the estimated phase.
This metric appropriately handles the circular nature of phase measurements while providing intuitive error interpretation: $\varepsilon = 0$ indicates perfect phase estimation, while $\varepsilon = \pi$ represents maximum error.

\subsection{Untethered mode}
\label{sec:results_untethered}

\begin{figure*}[t]
  \centering
  \hfill
  \subfloat[]{%
    \includegraphics[scale=0.55]{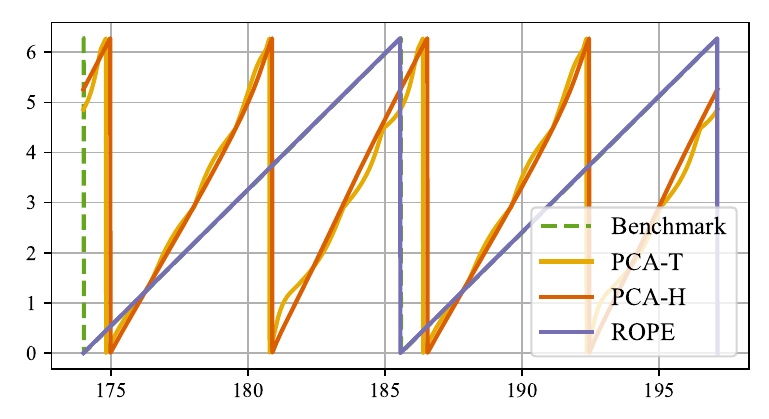}%
    \label{fig:figure_rossler_phase}
  }\hfill
  \subfloat[]{%
    \includegraphics[scale=0.55]{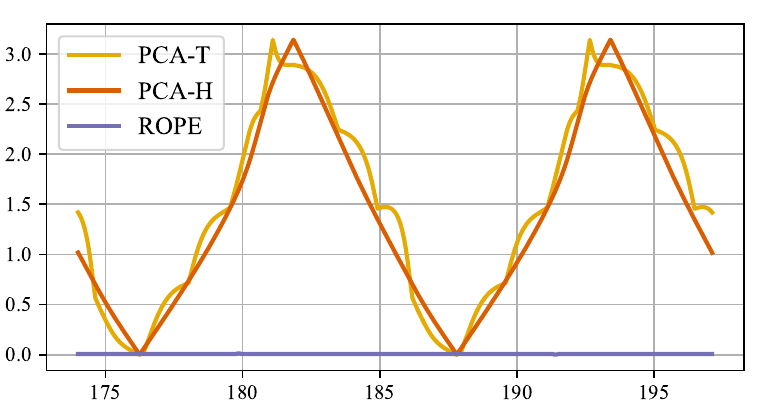}%
    \label{fig:figure_rossler_error}
  }\hfill~\\
  \hfill
   \subfloat[]{%
    \includegraphics[scale=0.55]{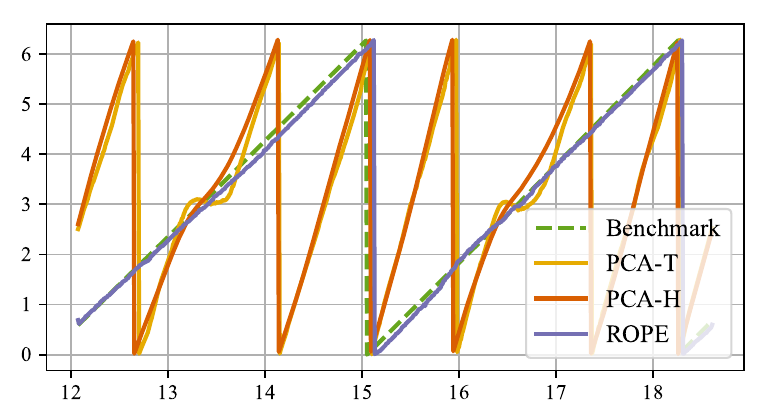}%
    \label{fig:figure_spiral_phase}
  }\hfill
  \subfloat[]{%
    \includegraphics[scale=0.55]{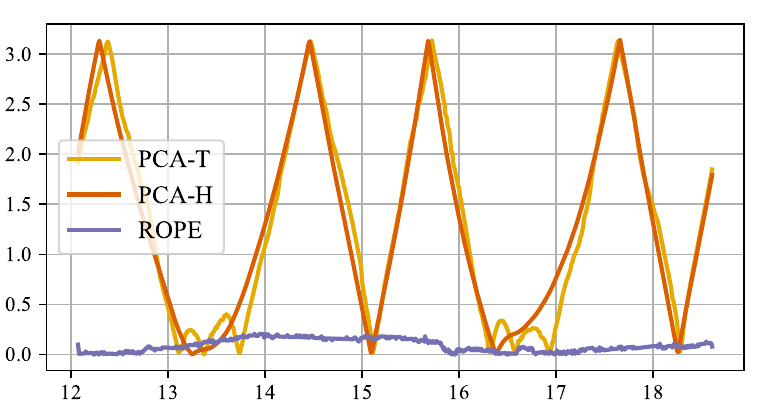}%
    \label{fig:figure_spiral_error}
  }\hfill~\\  
  \hfill
  \subfloat[]{%
    \includegraphics[scale=0.55]{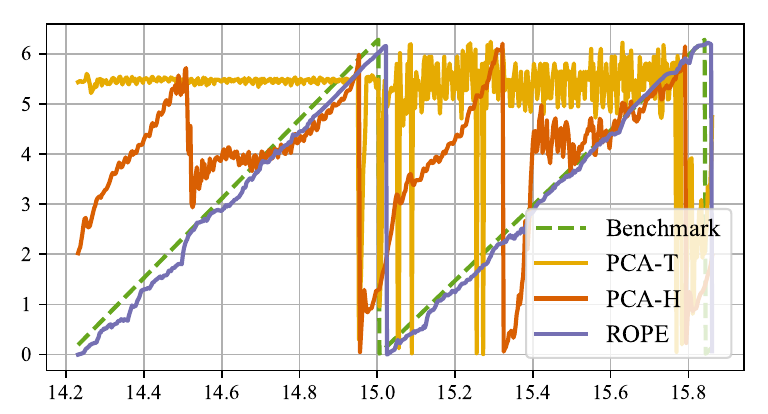}%
    \label{fig:figure_heart_phase}
  }\hfill
  \subfloat[]{%
    \includegraphics[scale=0.55]{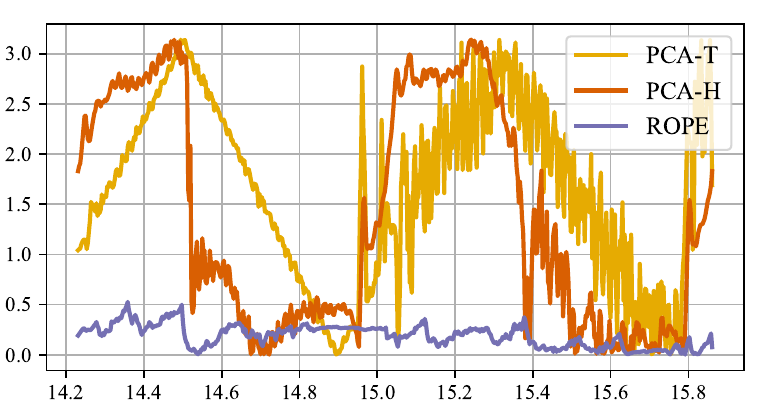}%
    \label{fig:figure_heart_error}
  }\hfill~
  \caption{Estimated phase trajectories $\theta$ (left column) and corresponding phase estimation errors $\varepsilon$ (right column) for representative signals from the validation dataset. 
  Each row shows a different signal type: (a,b) Rössler oscillator, (c,d) Spiral motion, (e,f) ECG signal. Two benchmark pseudo-periods are shown in each plot.}
  \label{fig:figure_multifig_phase}
\end{figure*}

\begin{figure*}[t]
  \centering
  \subfloat[]{%
    \includegraphics[scale=0.455]{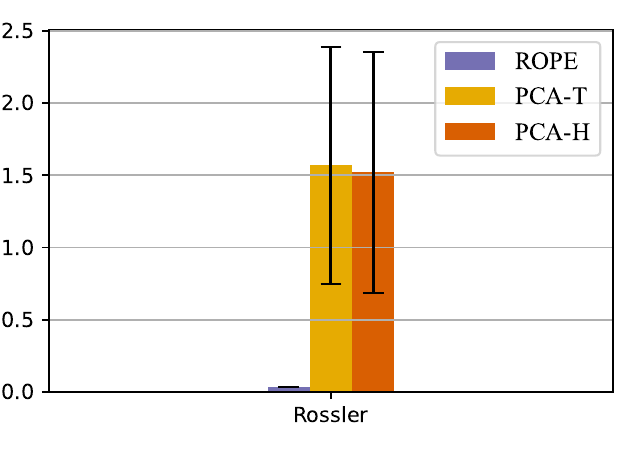}%
    \label{fig:bar_figure_rossler}
  }
  \hfill
  \subfloat[]{%
    \includegraphics[scale=0.45]{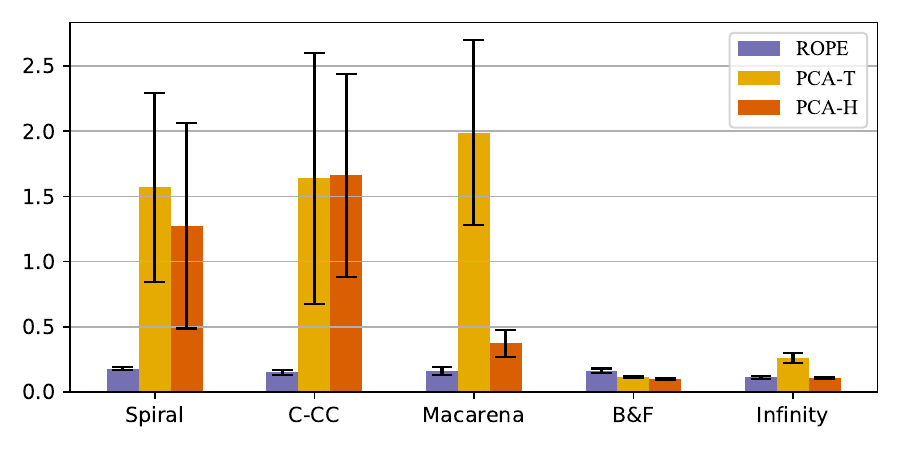}%
    \label{fig:bar_figure}
  }
  \hfill
  \subfloat[]{%
    \includegraphics[scale=0.45]{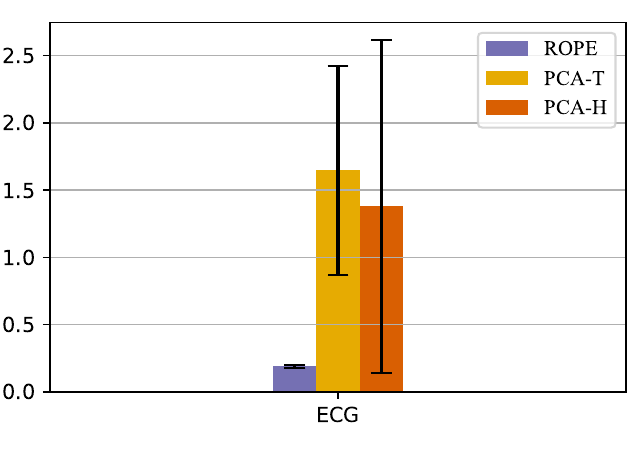}%
    \label{fig:bar_figure_ecg}
  }
  \caption{Mean and variance of phase estimation error $\varepsilon$ with respect to user defined benchmark phase for (a) Rössler oscillator, (b) human motion-capture signals, and (c) ECG.}
  \label{fig:bars}
\end{figure*}

Figure~\ref{fig:figure_multifig_phase} presents estimated phase trajectories (left column) and the corresponding phase estimation errors $\varepsilon$ (right column) for three representative signal types: a chaotic oscillator (Rössler), a human-generated motion trajectory (Spiral), and a physiological time series (ECG). 
In each scenario, the performance of the proposed ROPE algorithm is compared against those of the PCA-T and PCA-H algorithms, and evaluated with respect to the benchmark phase. 
All computations were performed offline on previously recorded trajectories and exploiting the accelerated computation of \eqref{eq:minimization_search} reported in Appendix \ref{sec:accelerated_minimization_search}.
In each case, two full pseudo-periods are shown.
Comprehensive quantitative performance metrics, including statistical summaries and variance analysis across all tested conditions, are presented in  Figure~\ref{fig:bars} and Table~\ref{tab:mean_variance}. 

When estimating the phase of the Rössler oscillator (Figures \ref{fig:figure_rossler_phase}, \ref{fig:figure_rossler_error})
we observe almost perfect estimation by the ROPE algorithm, with a phase that progresses linearly between one delimiter time instant and the next, without irregularities.
In contrast, PCA-T and PCA-H estimate more irregular phase curves and, crucially, are unable to correctly identify the start and end of the pseudo-periods.
The cause of this is the use of the PCA to reduce dimensionality of the signal, yielding fatal loss of information.
The same issue also occurs when estimating the phase of the Spiral motion (Figures~\ref{fig:figure_spiral_phase}, \ref{fig:figure_spiral_error}), where both PCA-T and PCA-H incorrectly detect three distinct pseudo-periods within each actual benchmark pseudio-period. 
This behavior is due to the spatial geometry of the trajectory, which contains three visible circles (see Figure \ref{fig:shapes}, Spiral).
%
The superior performance of the ROPE algorithm is generally confirmed also for the other human motion signals portrayed in Figure \ref{fig:shapes}.
As a matter of fact, the estimation error is significantly smaller both in mean and variance for more complex motions (Spiral, Macarena, Clockwise-Counterclockwise), and on par with PCA-H and PCA-T on simpler motions (Back\&Forth and Infinity), which are for this suited to PCA-based methods.
ROPE performs significantly better than PCA-T and PCA-H also when estimating the phase of the ECG recordings (Figures 
\ref{fig:ECG_shapes}).
Interestingly, unlike the previous datasets, ECG signals are intrinsically one-dimensional and do not require dimensionality reduction, which is skipped in this case when applying PCA-T and PCA-H.
Nevertheless, both display limited reliability in this setting, whereas ROPE shows consistently low error and adherence to the benchmark phase; see Figures \ref{fig:figure_heart_phase}, \ref{fig:figure_heart_error}, and Table \ref{tab:mean_variance} (ECG).

\begin{table*}[t]
    \centering
    \begin{tabular}{lcccccc}
        \toprule
        \textbf{Signal} & \multicolumn{6}{c}{\textbf{Error metrics}}\vspace{1ex}\\
         & \multicolumn{2}{c}{\emph{ROPE}} & \multicolumn{2}{c}{\emph{PCA-T}} & \multicolumn{2}{c}{\emph{PCA-H}} \\
        & $\mu$ & $\sigma^2$ & $\mu$ & $\sigma^2$ & $\mu$ & $\sigma^2$ \\
        \midrule
        Spiral    & 0.180 & 0.013 & 1.569 & 0.728 & 1.276 & 0.789 \\
        C-CC      & 0.150 & 0.018 & 1.639 & 0.964 & 1.661 & 0.777 \\
        Macarena  & 0.163 & 0.029 & 1.990 & 0.710 & 0.374 & 0.104 \\
        Infinity  & 0.111 & 0.010 & 0.259 & 0.040 & 0.106 & 0.007 \\
        B\&F      & 0.163    & 0.010    & 0.259    & 0.040    & 0.106    & 0.007    \\
        Rössler   & 0.032 & 0.001 & 1.562 & 0.797 & 1.520 & 0.833 \\
        ECG & 0.179 & 0.006 & 1.552 & 0.939 & 1.394 & 1.285 \\
        \bottomrule
    \end{tabular}
    \caption{Mean ($\mu$) and variance ($\sigma^2$) of the phase estimation error $\varepsilon$ with respect to the benchmark phase for each phase estimation method and signal type. 
    }
    \label{tab:mean_variance}
\end{table*}

\subsection{Tethered mode}

\begin{figure*}[t]
  \centering
  \subfloat[]{%
    \includegraphics[scale=0.55]{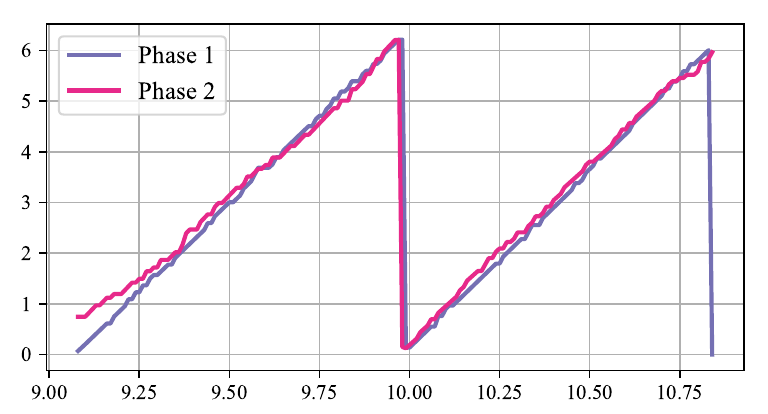}%
    \label{fig:antiphase_off}
  }
  \hspace{1.5cm} 
  \subfloat[]{%
    \includegraphics[scale=0.55]{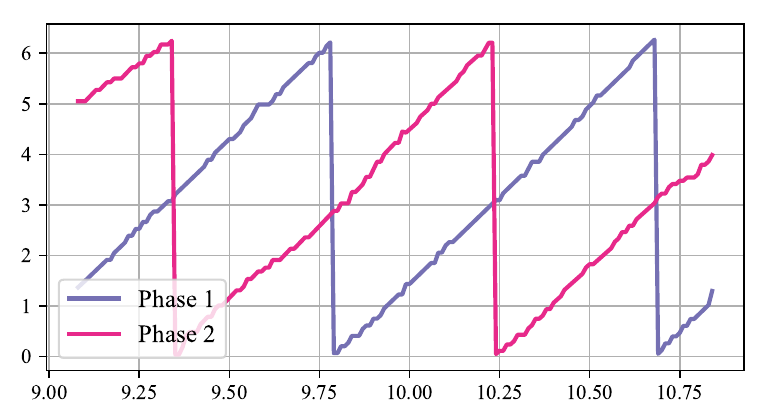}%
    \label{fig:antiphase_on}
  }\hfill
  \caption{Comparison of phase estimates in untethered and tethered mode during a motor synchronization task. (a) Untethered mode: the two signals (wrongly) appear phase-synchronized. (b) Tethered mode: the signals are correctly represented in anti-phase, enabled by proper offset alignment.}
  \label{fig:multifig_antiphase}
\end{figure*}

To validate the algorithm in tethered mode (cf.~Section \ref{Sec:algorithm}), we exploited $4$ $30$-second recordings where $2$ participants performed the Back \& Forth  motion (Figure \ref{fig:vicon_foto}), attempting to move in anti-phase (when the arm of a participant is flexed, the arm of the other is extended). 
Figure \ref{fig:antiphase_off} shows the phase estimation in untethered mode, while Figure \ref{fig:antiphase_on} displays the estimation in the tethered mode.
In the former case, the two signals appear to be in (in-phase) synchronization, which is not the case in reality, as phase zero must be associated to the same position for each participant, say with the arm flexed.
Conversely, in the tethered mode, the participant's signals are correctly assessed to be in anti-phase synchronization. 

\section{Conclusions}

In this work, we introduced ROPE (Recursive Online Phase Estimation), as a transformative approach to analyzing complex biological rhythms in real-time, addressing fundamental limitations that have constrained previous phase estimation methods when applied to authentic biological signals.
Unlike traditional techniques, ROPE does not rely on dimensionality reduction and it operates online making it suitable for real-time applications.
The algorithm's biological pattern-matching philosophy enables immediate deployment in demanding applications such as adaptive rehabilitation robotics, where precise timing of therapeutic interventions must align with individual movement patterns, and closed-loop medical devices, where treatment delivery depends on accurate real-time phase tracking of physiological rhythms. 

We established a rigorous mathematical foundation for analyzing pseudo-periodic biological signals---those exhibiting natural cycle-to-cycle variability characteristic of living systems---and developed a comprehensive algorithmic framework that operates effectively in multidimensional spaces, without the information loss inherent in traditional dimensionality reduction approaches.
We validated ROPE's capabilities through a systematic progression from controlled synthetic systems to challenging real-world biological signals: chaotic dynamics that model complex neural oscillations (Rössler system), natural human movement patterns exhibiting inherent motor variability, and clinical-grade electrocardiographic recordings containing authentic measurement noise and physiological variations.
Comprehensive comparison against established methods, including principal component analysis with Hilbert transform and real-time PCA-based approaches, revealed ROPE's superior accuracy and temporal consistency across all tested conditions.
Most significantly, ROPE maintained robust performance under challenging conditions that cause traditional methods to fail. These include multidimensional geometric complexity, nonlinear dynamics, and measurement noise that characterize authentic biological signals rather than idealized laboratory conditions.

Future developments will advance both the theoretical foundations and practical capabilities of biological phase estimation. Establishing formal error bounds that relate phase estimation accuracy to signal characteristics and pseudo-periodicity parameters will provide the theoretical guarantees needed for clinical deployment, enabling researchers and clinicians to predict algorithm performance for specific applications and patient populations. 
Equally important, enhancing the algorithm’s ability to automatically detect when the underlying motion pattern has fundamentally changed---such as when a subject transitions to a different type of movement---will significantly increase ROPE’s versatility. 
This capability would enable real-time monitoring of dynamic scenarios, such as adaptive rehabilitation sessions involving multiple exercises. 
\newpage
\appendix

\renewcommand{\thesection}{\alph{section}}

\section*{Appendix}

\addcontentsline{toc}{section}{Appendix} 

\section{Accelerated computation of \eqref{eq:minimization_search}}
\label{sec:accelerated_minimization_search}

To accelerate the computation of \eqref{eq:minimization_search}, we may limit the search domain (originally, $\C{P}_{i-1}$) to a window centered around the solution of \eqref{eq:minimization_search} at the previous time instant, that is $h^*(k-1)$.
Specifically, we define \emph{look-ahead} and \emph{look-behind intervals} $\delta^+, \delta^- \in \BB{N}_{>0}$, and restrict the search domain ($\C{P}_{i-1}$) in \eqref{eq:minimization_search} to (see Figure \ref{fig:search_interval})
\begin{equation}\label{eq:smaller_search_domain}
\begin{aligned}
    &\begin{dcases}
        \{\max\{\bar{k}_{i-1}, h^*(k-1) - \delta^-\}, \dots, h^*(k-1) + \delta^+\}, 
        & \text{if } h^*(k-1) + \delta^+ \le \bar{k}_i - 1,\\
        \begin{aligned}
        &\{\bar{k}_{i-1}, \dots, \bar{k}_{i-1} + \delta^+ - (\bar{k}_i - 1 - h^*(k-1))\} \\
        &\cup \{\max\{\bar{k}_{i-1}, h^*(k-1) - \delta^-\}, \dots, \bar{k}_i - 1\}
        \end{aligned}
        & \text{otherwise}.
    \end{dcases}
\end{aligned}
\end{equation}

\section{Relaxation of Assumption \ref{ass:signals}.\ref{ass:2}}
\label{sec:relaxation_ass_intersections}

When Assumption \ref{ass:signals}.\ref{ass:2} is not satisfied, the search for the minimum-distance point in $\C{P}_{i-1}$ in \eqref{eq:minimization_search} may return ambiguous results, as there can be multiple values of $h$ that minimize the argument of \eqref{eq:minimization_search}.
Here, we provide an extension to the algorithm presented in Section \ref{Sec:algorithm} that does not require Assumption \ref{ass:signals}.\ref{ass:2}.
Namely, note that $k - \bar{k}_i$ is the time elapsed since the start of the last pseudo-period $\C{P}_i$, at time $k \in \C{P}_i$; similarly, $h - \bar{k}_{i-1}$ is the time elapsed since the start of the previous pseudo-period $\C{P}_{i-1}$, at time $h \in \C{P}_{i-1}$; we measure a difference between these two quantities via the function%
\footnote{The second term in the minimization is introduced to have time instants ($k$) close to the end of pseudo-period $\C{P}_{i}$ being evaluated as proximal to initial time instants ($h$) in the previous pseudo-period $\C{P}_{i-1}$.}
\begin{equation}
    \begin{aligned}
        \sigma(h, k) \coloneqq {} & 
        \min \left\{
            \abs{h - \bar{k}_{i-1} - (k - \bar{k}_{i})},\ 
            \abs{h - \bar{k}_{i-1} + \kappa_{i-1} - (k - \bar{k}_{i})}
        \right\} \\
        = {} & \min \left\{
            \abs{h + \kappa_{i-1} - k},\ 
            \abs{h + 2\kappa_{i-1} - k}
        \right\}.
    \end{aligned}
\end{equation}
Then, we define the normalized function
$\sigma_{\R{n}}(h, k) \coloneqq \frac{\sigma(h, k)}{\max_{\hbar \in \C{P}_{i-1}} \sigma(\hbar, k)}$, and reformulate the minimization problem in \eqref{eq:minimization_search} as
\begin{equation}\label{eq:minimization_search_time}
  h^*(k) \coloneqq \arg\min_{h \in \C{P}_{i-1}} \left[
  \ON{dist}_\R{n}(\mathbf{P}_{i-1}, h, \vec{p}(k)) +
  \ON{dist}_\R{n}(\mathbf{V}_{i-1}, h, \vec{v}(k)) +
  \sigma_\R{n}(h, k)
  \right].
\end{equation}
With the reformulation in \eqref{eq:minimization_search_time}, the algorithm seeks to match the current signal state (at time $k$) to one in the previous period $\C{P}_{i-1}$ that not only has similar position and velocity, but also similar elapsed time since the start of the cycle.
Note that while this approach does not require Assumption \ref{ass:signals}.\ref{ass:2}, the estimation might become less accurate when $\varepsilon_\R{t}$ in Definition \ref{def:pseudoperiodicity} is relatively large, because the elapsed time becomes less relevant in identifying the current phase.

\enlargethispage{20pt}

\section*{Data Accessibility}
The ROPE phase estimation algorithm and the datasets used for validation are freely available at \url{https://github.com/SINCROgroup/recursive-online-phase-estimator}.

\section*{Authors' Contributions}
MC and MdB conceived the research. MdB, MC, and FDL supervised the research.
MC and AS formalized the mathematical problem. AS and MC designed and implemented the algorithm.
AS, FDL, and MC carried out the experiments. AS analyzed the data.
AS and MC wrote the initial draft of the manuscript.
All authors contributed to the interpretation of the results and edited the final manuscript.

\section*{Funding}
This work was in part supported by the Research Project “SHARESPACE” funded by the European Union (EU HORIZON-CL4-2022-HUMAN-01-14. SHARESPACE. GA 101092889).

\bibliographystyle{\bibliographyStyleName} 
\bibliography{\myBibliographyFile}

\end{document}

%% file: preamble_report.tex
\usepackage[utf8]{inputenc}
\usepackage[T1]{fontenc}
\usepackage{newtxtext}    
\usepackage{newtxmath}    

\usepackage{amsmath,amssymb,mathtools,bm}
\makeatletter
\@ifundefined{openbox}{}{}
\makeatother
\usepackage{amsthm}
\usepackage[svgnames]{xcolor}
\usepackage{graphicx}
\usepackage[caption=false,font=footnotesize]{subfig}
\usepackage{booktabs}
\usepackage[export]{adjustbox}
\usepackage[ruled,vlined,linesnumbered]{algorithm2e}
\usepackage{enumitem}
\usepackage{soul}
\usepackage{lipsum}
\usepackage{mdframed}
\usepackage{float}
\usepackage{geometry}
\usepackage{titlesec}
\usepackage{abstract}
\usepackage[hidelinks]{hyperref}
\usepackage{datetime}
\usepackage{fancyhdr}

\newcommand{\ON}[1]{\operatorname{#1}}
\newcommand{\BB}[1]{\mathbb{#1}}
\newcommand{\C}[1]{\mathcal{#1}}
\newcommand{\R}[1]{\mathrm{#1}}

\DeclarePairedDelimiter{\norm}{\lVert}{\rVert}
\newcommand{\T}{^{\top}} 

\newcommand{\abs}[1]{\left\lvert#1\right\rvert}

\providecommand{\myHeaderAuthorTitle}{} 

\ifnum\myBibliographyStyle=0       
	\usepackage[numbers]{natbib}
	\newcommand{\bibliographyStyleName}{IEEEtran}
\else\ifnum\myBibliographyStyle=1  
	\usepackage[square]{natbib}
	\newcommand{\bibliographyStyleName}{IEEEtranSN}
	\renewcommand{\cite}{\citep} 
\else
	\GenericError{}{Value of \protect\myBibliographyStyle\space is not acceptable}{}{}
\fi\fi

\graphicspath{{figures/}}         
\numberwithin{equation}{section}	
\allowdisplaybreaks               

\makeatletter
\patchcmd{\@algocf@start}
  {-1.5em}
  {0pt}
  {}{}
\makeatother

\SetKwComment{tcc}{ {$\triangleright$} }{}  
\SetKwComment{tcp}{ {$\triangleright$} }{}  

\SetCommentSty{myCommentStyle}

\ifnum\numColumns=1
\else\ifnum\numColumns=2
	\let\oldtabular\tabular
	\renewcommand{\tabular}{\footnotesize\oldtabular}
\else
  \GenericError{}{Value of \protect\numColumns\space is not acceptable}{}{}
\fi\fi
\setlist{noitemsep,topsep=0.5ex}     
\definecolor{myLightBlue}{rgb}{0.09 0.36 0.63}
\colorlet{colorAccent}{myLightBlue}
\colorlet{colorAccentLight}{gray!15}
\hypersetup{  
  colorlinks,
  linkcolor = {DodgerBlue},
  citecolor = {DodgerBlue},
  urlcolor  = {DodgerBlue}
}

\newdateformat{myDateFormat}{\THEDAY \ \monthname[\THEMONTH] \THEYEAR}
\let\oldheadrule\headrule
\let\oldfootrule\footrule
\renewcommand{\headrule}{{\color{colorAccent}\oldheadrule}}
\renewcommand{\footrule}{{\color{colorAccent}\oldfootrule}}
\fancypagestyle{firstpage}{  
	\fancyhf{}
	\lhead{\sffamily \footnotesize \myDocumentType}
	\chead{}
	\ifnum\dateMode=0
		\rhead{\sffamily \footnotesize}
	\else\ifnum\dateMode=1
		\rhead{\sffamily \footnotesize \myDateFormat\today}
	\else\ifnum\dateMode=2
		\rhead{\sffamily \footnotesize \myDate}
	\else
		\GenericError{}{Value of \protect\dateMode\space is not acceptable}{}{}
	\fi\fi\fi
	\lfoot{}
	\cfoot{\sffamily\footnotesize \thepage}
  \rfoot{}

}

\titleformat{\part}[display]{\centering\sffamily\Huge\bfseries}{\textsc{\partname}\thepart\\ \vspace*{1ex}{\Huge\color{colorAccent}------ $\cdot$ ------}}{1.2ex}{\huge}
\titleformat{\chapter}{\sffamily\bfseries\Huge\color{colorAccent}}{\thechapter \	}{10pt}{}
\ifnum\isUseThightHeaders=1
	\titleformat*{\section}{\centering\large\sffamily\color{colorAccent}}
	\titleformat*{\subsection}{\sffamily\color{colorAccent}}
	\titlespacing\section{0pt}{3ex}{1ex}
	\titlespacing\subsection{0pt}{2.5ex}{0.5ex}
\else
	\titleformat*{\section}{\Large\sffamily\color{colorAccent}}
	\titleformat*{\subsection}{\large\sffamily\color{colorAccent}}
\fi
\titleformat*{\subsubsection}{\sffamily\itshape\color{colorAccent}}
\titleformat*{\paragraph}{\color{colorAccent} \itshape \sffamily}
\titlespacing*{\paragraph}{0pt}{0pt}{10pt}
\titleformat*{\subparagraph}{\itshape}
\titlespacing*{\subparagraph}{0pt}{0pt}{10pt}

\newtheoremstyle{mythmstyle}
{0.7em}
{1em}
{\itshape}
{}
{\sffamily \color{colorAccent}}
{.}
{ }
{{\bfseries\thmname{#1}\thmnumber{ #2}}\thmnote{ (#3)}}  

\theoremstyle{mythmstyle}			 

\mdfdefinestyle{myshadedthm}{
	backgroundcolor  = colorAccentLight,
	linewidth        = 0pt,
	innerleftmargin  = 2ex,
	innerrightmargin = 2ex,
	innertopmargin   = 0ex
	}

\newmdtheoremenv[style=myshadedthm]{theorem}{Theorem}[section]

\newmdtheoremenv[style=myshadedthm]{corollary}{Corollary}[section]
\newmdtheoremenv[style=myshadedthm]{definition}{Definition}[section]
\newmdtheoremenv[style=myshadedthm]{assumption}{Assumption}[section]
\newmdtheoremenv[style=myshadedthm]{problem}{Problem}[section]
\newmdtheoremenv[style=myshadedthm]{remark}{Remark}[section]
\newmdtheoremenv[style=myshadedthm]{conjecture}[theorem]{Conjecture}

\newcommand{\B}[1]{\if#1\relax\bm{#1}\else\mathbf{#1}\fi} 

\renewcommand{\vec}[1]{\if#1\relax\bm{#1}\else\mathbf{#1}\fi}
\newcommand{\mat}[1]{\if#1\relax\bm{#1}\else\mathbf{#1}\fi}

\newcommand{\styleAuthor}{\color{colorAccent}\normalsize \sffamily \itshape}
\newcommand{\styleAffiliation}{\color{colorAccent}\scriptsize \sffamily \itshape}

\title{\color{colorAccent}{\Huge \sffamily \myTitle}}%
\ifnum\isUseThightHeaders=1
\fi
\author{%
  \begin{minipage}{\textwidth}\centering
    \myAuthor\\[0.6ex]
    {\styleAffiliation \myAffiliation}
  \end{minipage}%
  \ifnum\isUseThanks=1 \thanks{\myThanks\/}\fi
}

\date{}
\renewcommand\footnotemark{}

\makeatletter
\ifx\@date\empty
\patchcmd{\@maketitle}{{\@bspredate \@date \@bspostdate} \maketitlehookd \par \vskip 1.5em}{\vskip 0.5em}{}{}
\fi
\makeatother